\documentclass[usenatbib]{mnras}
\usepackage{newtxtext,newtxmath}

\usepackage[T1]{fontenc}

\DeclareRobustCommand{\VAN}[3]{#2}
\let\VANthebibliography\thebibliography
\def\thebibliography{\DeclareRobustCommand{\VAN}[3]{##3}\VANthebibliography}

\usepackage{graphicx}	
\usepackage{amsmath}	
\usepackage{caption}
\usepackage{subcaption}
\usepackage{booktabs}
\usepackage{makecell}
\usepackage{soul}
\usepackage{color}

\title[Detecting Galaxy Mergers: A Comparison]{Galaxy And Mass Assembly (GAMA): Comparing Visually and Spectroscopically Identified Galaxy Merger Samples}

\author[A. Desmons et al.]{
Alice Desmons,$^{1}$\thanks{E-mail: a.desmons@unsw.edu.au}
Sarah Brough,$^{1}$
Cristina Mart\'inez-Lombilla,$^{1}$
Roberto De Propris,$^{2,3}$
\newauthor
Benne Holwerda,$^{4}$
\'Angel R. L\'opez S\'anchez$^{5,6,7}$
\\
$^{1}$School of Physics, University of New South Wales, NSW 2052, Australia\\
$^{2}$Finnish Centre for Astronomy with ESO, University of Turku, Vesilinnantie 5, 20014, Turku, Finland\\
$^{3}$Department of Physics and Astronomy, Botswana International University of Science and Technology, Private Bag 16, Palapye, Botswana\\
$^{4}$University of Louisville, Department of Physics and Astronomy, 102 Natural Science Building, 40292 KY Louisville, USA\\
$^{5}$Australian Astronomical Optics, Macquarie University, 105 Delhi Rd, North Ryde, NSW 2113, Australia\\
$^{6}$Macquarie University Research Centre for Astronomy, Astrophysics \& Astrophotonics, Sydney, NSW 2109, Australia\\
$^{7}$ARC Centre of Excellence for All Sky Astrophysics in 3 Dimensions (ASTRO-3D), Australia\\
}
\date{Accepted 2023 May 26. Received 2023 May 21; in original form 2023 February 7}

\pubyear{2023}

\begin{document}
\label{firstpage}
\pagerange{\pageref{firstpage}--\pageref{lastpage}}
\maketitle

\begin{abstract}
We conduct a comparison of the merging galaxy populations detected by a sample of visual identification of tidal features around galaxies as well as spectroscopically-detected close pairs of galaxies to determine whether our method of selecting merging galaxies biases our understanding of galaxy interactions. Our volume-limited parent sample consists of 852 galaxies from the Galaxy And Mass Assembly (GAMA) survey in the redshift range $0.04~\leq~z~\leq~0.20$ and stellar mass range $9.50~\leq~$log$_{10}(M_{\star}/\rm{M}_{\odot})\leq~11.0$. We conduct our comparison using images from the Ultradeep layer of the Hyper Suprime-Cam Subaru Strategic Program (HSC-SSP) to visually-classify galaxies with tidal features and compare these to the galaxies in the GAMA spectroscopic close-pair sample. We identify 198 galaxies possessing tidal features, resulting in a tidal feature fraction $f_{\rm{tidal}}$~=~0.23~$\pm$~0.02. We also identify 80 galaxies involved in close pairs, resulting in a close pair fraction $f_{\rm{pair}}$ = 0.09~$\pm$~0.01. Upon comparison of our tidal feature and close pair samples we identify 42 galaxies that are present in both samples, yielding a fraction $f_{\rm{both}}$~=~0.05~$\pm$~0.01. We find evidence to suggest that the sample of close pairs of galaxies is more likely to detect early-stage mergers, where two separate galaxies are still visible, and the tidal feature sample detects later-stage mergers, where only one galaxy nucleus remains visible. The overlap of the close pair and tidal feature samples likely detect intermediate-stage mergers. Our results are in good agreement with the predictions of cosmological hydrodynamical simulations regarding the populations of merging galaxies detected by close pair and tidal feature samples. 
\end{abstract}




\section{Introduction}
\label{sec:intro}

The currently accepted model for the evolution of galaxies in the early Universe into the galaxy populations we observe today is known as the “Hierarchical Merger Model”. This model is a direct consequence of the $\Lambda$CDM model of the Universe, and postulates that the growth of the universe's highest-mass galaxies is dominated by merging with lower-mass galaxies (e.g. \citealt{Lacey1994NBodyMergeRate,Cole2000Hierarchical,Robotham2014GAMAClosePair, Martin2018MergeMorphTransform}). Observationaly we have been able to study the impact of major mergers (mass ratios > 1:4) on galaxy evolution using close pairs of galaxies (e.g. \citealt{DePropris2005MillenniumPairs,Holwerda2011HIMorph,Robotham2014GAMAClosePair, Banks2021GAMAMergingBCG}). However, minor mergers are thought to play a significant role in galaxy evolution. For example \citet{Martin2018MergeMorphTransform} use the Horizon-AGN cosmological hydrodynamical simulation to study the morphological evolution of galaxies with stellar masses above $10^{10}$ M$_{\odot}$. They show that on average, elliptical galaxies have undergone more mergers than disc-like galaxies and that minor mergers (mass ratios < 1:3) are responsible for the majority of galaxy morphological transformation at $z<1$.

Although spectroscopic samples of close pairs are crucial to advancing our knowledge of the role of mergers in galaxy evolution, they possess some limitations. If a secondary galaxy has already been ripped apart or absorbed into its host galaxy or if the secondary galaxy is too low-mass to detect it will not be included in the sample. Spectra are also often only measured for brighter (i.e. more massive) galaxies than those detected in imaging. This means that close-pair samples will be biased towards brighter, more massive merging systems and will not provide information about mergers with the more prevalent low-mass galaxies. As a consequence, we lack observational confirmation on the contribution of low-mass galaxies to the merging process.

Galaxies undergoing mergers typically also show signs of disturbance in their morphology, known as tidal features. These tidal features present as faint, diffuse, non-uniform regions of stars that extend into space from a galaxy, and can only be detected in very deep images (e.g. \citealt{LopezSanchez2010WolfRayetStarForm}). By using tidal features to detect merging galaxies, we can detect fainter mergers and mergers where the secondary galaxy is too low-mass to be detected by close pair samples, allowing us to assemble a more complete sample of merging galaxies and better characterise the impact of mergers on galaxy evolution \citep{Johnston2008GalFormStellarHalo}.

Detection of tidal features is typically achieved through visual identification of optical images. Tidal features are classified based on their morphology (e.g. \citealt{Atkinson2013CFHTLSTidal,KadoFong2018HSCTidalFeat, Bilek2020MATLASTidalFeat}). They are generally separated into two main classes: "shells", which are umbrella-like envelopes of stars curving around the host galaxy, often forming concentric arcs (e.g. \citealt{Malin1983EllipGalShells,Quinn1984ShellsEllipGals}), and "streams", which are narrow filaments that emanate from the primary galaxy (e.g. \citealt{Toomre1972BridgeTails,Kaviraj2019ADNDwarfGal}). The type of tidal features found around a galaxy depends on the properties of the host galaxy. The properties of these features can reveal information about the merger which created them, such as the mass of the progenitor galaxies and the morphology of the galaxies involved in the merger, either elliptical or disk-like \citep{Hendel2015TidalDebOrbit}.

Past research involving the visual identification and classification of tidal features has mainly been based on observations of relatively small samples of galaxies (see, e.g. \citealt{MartinezDelgado2009NGC4013TidalStream,Holwerda2011HIMorph,MartinezDelgado2015NGC4631TidalStream, Hood2018RESOLVETidalFeat, Bilek2020MATLASTidalFeat}). These generally define a tidal feature fraction, where $f_{\rm{tidal}}$ is the fraction of galaxies in a given sample that host tidal features. In these works, $f_{\rm{tidal}}$ typically ranges from 0.10~$\leq~f_{\rm{tidal}}~\leq$~0.30 depending on the selection criteria of the parent sample (e.g. parent survey, galaxy type, environment) and whether the authors included only distinct tidal features (e.g. \citealt{Hood2018RESOLVETidalFeat}) or also included weaker, potential tidal features (e.g. \citealt{Atkinson2013CFHTLSTidal}) in their sample.

There also exist other methods of detecting merging galaxies, for example using optical coefficients such as concentration, asymmetry, Gini, or $M_{20}$ (e.g. \citealt{Conselice2003LightDistGalForm,Conselice2008MergerRate,Conselice2009MergerHist,Lotz2011MajorMinorRate}). One can also study the distribution and velocity structure of the neutral hydrogen gas in galaxies to detect disturbed and merging galaxies (e.g. \citealt{Holwerda2011HIMorph}). Although these methods are also useful ways of detecting interacting galaxies, they are not the focus of this paper.

The different methods of detecting interacting galaxies have also been considered through analysis of cosmological hydrodynamical and N-body simulations of galaxy evolution. One such simulation performed by \citet{Lotz2011MajorMinorRate} used hydrodynamical merger simulations to estimate the observability timescale (T$_{\rm{obs}}$) of ongoing mergers and found T$_{\rm{obs}}\sim$~600~Myr for close pairs with a range of projected separations. Tidal features, on the other hand, are estimated to remain observable for a few billion years after a merger both by $N$-body simulations (e.g. \citealt{Hendel2015TidalDebOrbit}) and observational studies (e.g. \citealt{Huang2022HSCTidalFeatETG}). The significantly longer observability period of tidal features compared to close pairs makes them a valuable tool in the detection of galaxy mergers, both past and present.

One caveat of using visual classification to identify merging galaxies is the introduction of human bias into the analysis. \citet{Martin2022TidalFeatMockIm} conducted an analysis of the frequency and visibility of tidal features using $\sim$~8000 mock images from the hydrodynamical simulation \texttt{NEWHORIZON} \citep{Dubois2021NewHorizon} classified by 45 expert classifiers. They also obtained a measure of variation between classifiers using a subset of $\sim$~600 images which were each classified 5 times. \citet{Martin2022TidalFeatMockIm} found that below a limiting surface brightness $\mu_{r}\sim$~30~mag~arcsec$^{-2}$, galaxy projection (or viewing angle) was the more significant source of disagreement between classifications rather than disagreement between individual classifiers. They also found that while concurrence between classifiers improved with deeper imaging, galaxy morphologies became more complex, which introduced more uncertainty to the precise characterisation of the tidal features.

Recent works have attempted to overcome such issues by using machine learning techniques to detect and classify merging galaxies. Machine learning models relying on labelled training sets from observational surveys have shown great success (e.g. \citealt{Pearson2019DeepLearnMergers, Walmsley2019CNNTidalFeat}) but potentially learn the same bias introduced by visual classification as the galaxies used for training have to be identified visually. However, works using machine learning models trained on images from hydrodynamical simulations (e.g. \citealt{Snyder2019IllustrisAutoMergerClass, Bottrell2019DeepLearningMerger}) to classify merging galaxies have access to the full history of simulated galaxies and are therefore confident that a galaxy has been involved in a recent merger. \citet{Bottrell2019DeepLearningMerger} use idealised and `fully real' images (simulated images inserted into real SDSS survey fields) from hydrodynamical simulations to investigate how machine learning models perform when trained on different types of simulated images. They found that even networks trained on the `fully real' images reached high (87\%) accuracy, suggesting that training models on simulated data before applying them to real datasets may be a way of automating, and removing observational bias from tidal feature detection.

Our goal in this work is to understand the differences in the galaxies selected based on visually-identified tidal features compared to spectroscopic close pairs. We use the Galaxy And Mass Assembly (GAMA, \citealt{Driver2011GAMADataRel}) survey and its associated spectroscopically-identified Close Pair sample \citep{Robotham2011GAMAGroups,Robotham2014GAMAClosePair}. The Hyper Suprime-Cam Subaru Strategic Program (HSC-SSP; \citealt{Aihara2018HSCSurveyDesign}) survey Ultradeep layer provides galaxy images with the sensitivity required to detect faint tidal features, with a limiting depth of $\mu_{r}\sim$ 28 mag arcsec$^{-2}$. These images, paired with the GAMA spectroscopic catalogues which provide accurate distances to the HSC-SSP galaxies, provide the first opportunity to compare an observationally detected visually-identified tidal feature sample and spectroscopic close pair sample for the same dataset. The primary aim of this work is to determine whether our method of selecting merging galaxies biases our understanding of galaxy interactions.

Section \ref{sec:data} details the data sources used in this work and the selection of our sample of galaxies. Section \ref{sec:results} presents the results of our analysis including the fractions of galaxies in close pairs and galaxies with tidal features as well as the dependence of these fractions on galaxy properties. Section \ref{sec:disc} compares the results obtained in this paper to previous observational studies, and models. We present our conclusions in Section \ref{sec:conc}.
Throughout this paper, we assume a flat $\Lambda$CDM cosmology with $h$ = 0.7, $H_{0}$ = 100~$h$ km s$^{-1}$ Mpc$^{-1}$, $\Omega_{m}$~=~0.3, and $\Omega_{\Lambda}$ = 0.7.

\section{Data Sources and Sample Selection}
\label{sec:data}

The data for this work is sourced from the GAMA survey for spectroscopic data and the HSC-SSP Public Data Release 2 (PDR2) for deep galaxy images. In this Section we detail the data we use; the process of selecting a volume-limited sample; how the classification of this sample will be performed and the assignment of confidence levels to these classifications; how we assign a colour cut to classify red, more likely to be early-type, and blue, more likely to be late-type, galaxies; and finally how we calculated the uncertainties presented in the analysis of our sample.

\subsection{The Galaxy And Mass Assembly survey (GAMA)}
\label{sec:gama} 

The GAMA survey is a spectroscopic survey carried out at the 3.9m Anglo-Australian Telescope (AAT) over 210 nights (2008-2014) using the optical AAOmega multi-object spectrograph \citep{Driver2011GAMADataRel,Hopkins2013GAMASpectra,Liske2015GAMADR2,Baldry2018GAMAG02,Driver2022GAMADR4}. The survey is split into five regions covering $\sim$~286~deg$^2$ down to a depth of \textit{r} < 19.8 mag and consists of $\sim$~300,000 galaxies. For this work, we focus on the GAMA G02 region delimited by 30.2$^{\circ}\leq$ R.A. $\leq$ 38.8$^{\circ}$ and -10.25$^{\circ}\leq$ Decl. $\leq$ -3.72$^{\circ}$, as this is where the HSC-SSP PDR2 \citep{Aihara2019HSCSecondData} Ultradeep images overlap the GAMA sample. The G02 region of the GAMA survey has a high spectroscopic completeness of 95.5\%, achieved through multiple visits to the same fields \citep{Baldry2018GAMAG02}. This makes it an excellent resource for this work. 

The GAMA survey data is stored in a database which is separated in Data Management Units (DMUs) that combine specific data to be used in different analyses. In this paper we make use of the GAMA Galaxy Group Catalogue (G$^{3}$C, \citealt{Robotham2011GAMAGroups}), which provides positions and redshifts for GAMA galaxies used to select the GAMA Close Pair sample, and the \texttt{StellarMasses} DMU to obtain galaxy stellar masses.

\subsubsection{GAMA G$^{3}$C}
\label{sec:groupfinding}

The G$^{3}$C \citep{Robotham2011GAMAGroups} contains four data tables, two of which will be used for this project, namely the \texttt{G3CGalv10} and \texttt{G3CGalPairv10} catalogues. The \texttt{G3CGalv10} catalogue contains properties of 204,110 galaxies from the parent GAMA sample which were selected by setting a limit of the redshift quality and imposing restrictions on the CMB-frame redshift $0.003 < z_{\rm{CMB}} < 0.6$. Additionally, to be included in the \texttt{G3CGalv10} catalogue, galaxies had to have a Survey Class ranking greater than four, meaning they have r-band magnitude $r < 19.8$ mag and satisfy \textit{r}-band star-galaxy separation. These galaxies were then used to select the GAMA close pair sample.
We use a subset of the G02 region data since we are only interested in the region overlapping with the Ultradeep HSC-SSP region. Hence we only keep galaxies in the central portion of the Ultradeep HSC-SSP observations, shown by the blue rectangle in Figure \ref{fig:HSC_RA_DEC}, for which 34.8$^{\circ}$< R.A.< 36.5$^{\circ}$ and -5.4$^{\circ}$< Decl.< -4.0$^{\circ}$. Applying these cuts, we are left with 2327 galaxies in the region where the GAMA G02 region overlaps the Ultradeep HSC-SSP region.

The second catalogue we make use of is the \texttt{G3CGalPairv10} catalogue, which contains information about the GAMA close pair sample. The GAMA close pair sample uses a friends-of-friends (FoF) algorithm to classify galaxy pairs \citep{Robotham2011GAMAGroups, Robotham2014GAMAClosePair} using three thresholds of projected spatial separation $r_{\rm{sep}}$ and radial velocity separation $v_{\rm{sep}}$, namely:
\begin{itemize}
    \item $r_{\rm{sep}}< 20~h^{-1 }$ kpc $\wedge$ $v_{\rm{sep}} < 500$ km s$^{-1}$
    \item $r_{\rm{sep}}< 50~h^{-1 }$ kpc $\wedge$ $v_{\rm{sep}} < 500$ km s$^{-1}$
    \item $r_{\rm{sep}}< 100~h^{-1 }$ kpc $\wedge$ $v_{\rm{sep}} < 1000$ km s$^{-1}$
\end{itemize}
To be confident that the close pairs we select are definitely interacting we use the second threshold and only select close pairs within 50 \textit{h}$^{-1}$ kpc and 500 km s$^{-1}$ of each other. The full close pair catalogue contains 31,823 galaxy pairs, of which 11,213 satisfy our $r_{\rm{sep}}$ and $v_{\rm{sep}}$ threshold. After applying our Right Ascension and Declination cuts, we are left with 125 galaxy pairs in the region where the GAMA G02 region overlaps the Ultradeep HSC-SSP region.

\subsubsection{Stellar Masses}
\label{sec:stellarmass}

We use the \texttt{StellarMasses} DMU \citep{Taylor2011GAMAMassEst} to obtain galaxy stellar masses. These are measured by stellar population synthesis (SPS) modelling. For the G02 region the masses are estimated using photometry from the Canada–France–Hawaii Telescope Legacy Survey (CFHTLS; \citealt{Ilbert2006CFHTLSPhotoZ}) which is preferred as it is deeper. When that is not available, we use masses estimated using photometry from the Sloan Digital Sky Survey (SDSS; \citealt{York2000SDSS}). For our sample of 2327 galaxies, 97\% of the stellar masses come from CFHTLS, 3\% come from SDSS, and 2 galaxies do not have mass information, reducing our parent GAMA sample to 2325 galaxies.

\subsection{The Hyper Suprime Cam Subaru Strategic Program (HSC-SSP)}
\label{sec:hsc}

\begin{figure}
    \centering
	\includegraphics[width=0.8\columnwidth]{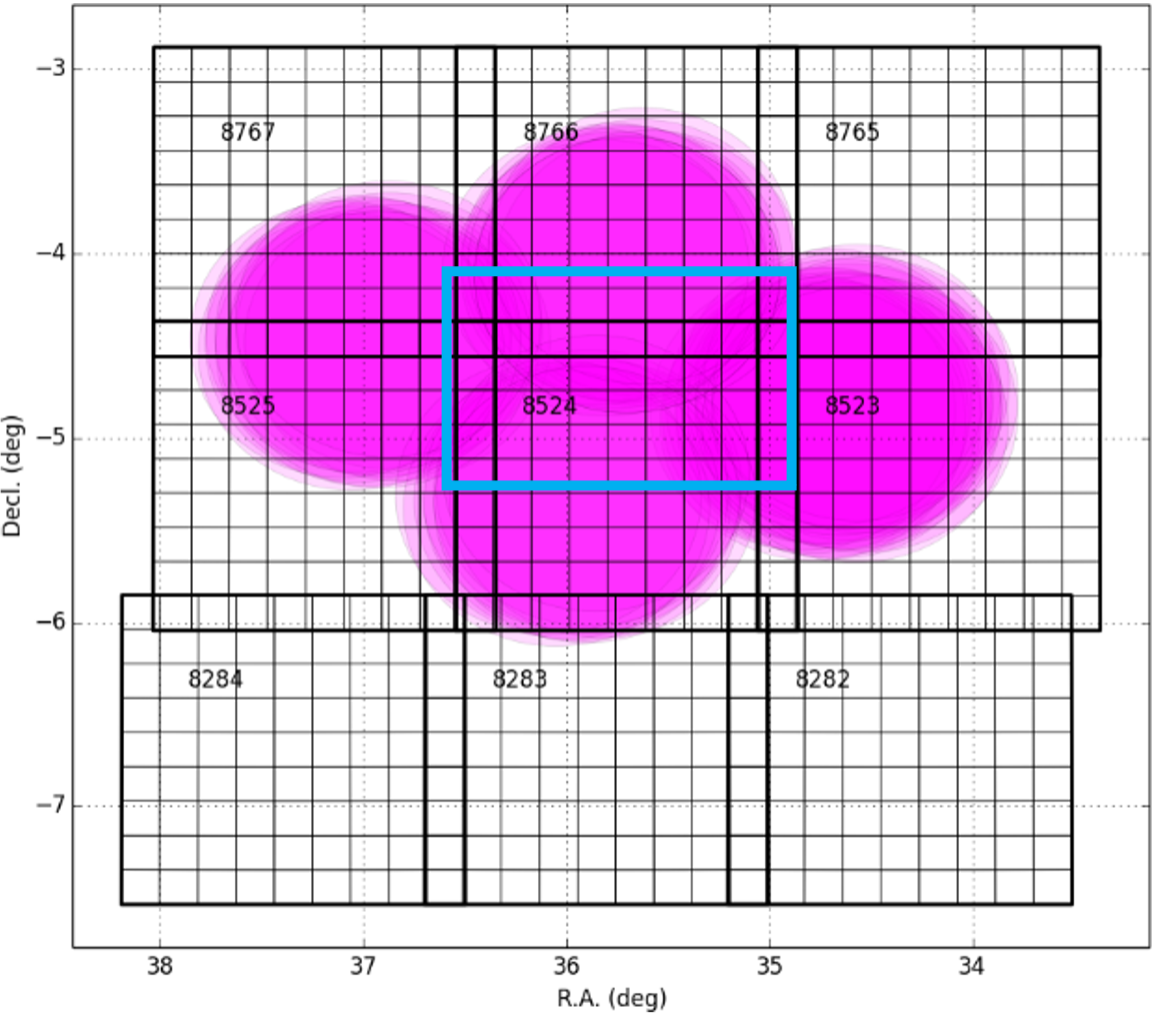}
    \caption{The pink region shows the Ultradeep layer of the HSC-SSP survey. The blue rectangle shows the sample selection applied here. Adapted from the HSC-SSP Available Data website \protect\footnotemark.}
    \label{fig:HSC_RA_DEC}
\end{figure}

The HSC-SSP survey \citep{Aihara2018HSCSurveyDesign} is a three-layered, \textit{grizy}-band imaging survey carried out with the Hyper Suprime-Cam (HSC) on the 8.2m Subaru Telescope located in Hawaii. In this work we use the images from the HSC-SSP Public Data Release 2 (PDR2; \citealt{Aihara2019HSCSecondData}). During the development of this project the HSC Public Data Release 3 (HSC-PDR3; \citealt{Aihara2022HSCThirdData}) became available. Although this is an updated version of the HSC-PDR2 we do not use this data due to the differences in the data treatment pipelines. HSC-PDR2 fulfils the requirements for our study and has been widely tested for low surface brightness studies (e.g. \citealt{Huang2018HSCStellarHalo,Huang2020WeakLens,Li2022MassiveGalOutskirt,MartinezLombilla2023GAMAIntragroupLight}).
The survey comprises of three layers: Wide, Deep, and Ultradeep which are observed to surface brightness depths of $\mu_{r}\sim$~26.4, $\mu_{r}\sim$~27.4, and $\mu_{r}\sim$ 28.0~mag arcsec$^{-2}$ respectively.
We use the Ultradeep region pictured in Figure \ref{fig:HSC_RA_DEC}, with the blue rectangle showing our sample's right ascension and declination limits. We use this region as it is the only HSC-SSP PDR2 Ultradeep region that overlaps with GAMA. The HSC-SSP PDR2 has a median $\it{i}$-band seeing of 0.6 arcsec and a resolution of 0.168 arcsec per pixel.

\footnotetext{\url{https://hsc-release.mtk.nao.ac.jp/doc/index.php/data-2/}}

To access the HSC-SSP galaxy images, we use the `Unagi' Python tool \citep{Huang2019Unagi} which allows us to create multi-band ‘HSC cutout’ images of size 355 x 355 pixels or 60 x 60 arcsecs.

Our analysis of the tidal feature and close pair samples includes a comparison of \textit{g – i} galaxy colour, as a general indicator of galaxy morphology, between the two samples. We retrieve the rest-frame \textit{g-} and \textit{i-}band magnitudes for the galaxies in our sample from the HSC-SSP PDR2  \texttt{photoz$\_$mizuki} catalogue as these results are deeper than the CFHT or SDSS measurements used for the GAMA stellar mass estimates described in Section \ref{sec:stellarmass}. These rest-frame magnitudes were inferred from the results of the \texttt{MIZUKI} \citep{Tanaka2015PhotoZBayesionPrior} template-fitting code. 

\subsection{Sample Selection}
\label{sec:sampleselect}

\begin{figure}
    \centering
	\includegraphics[width=0.95\columnwidth]{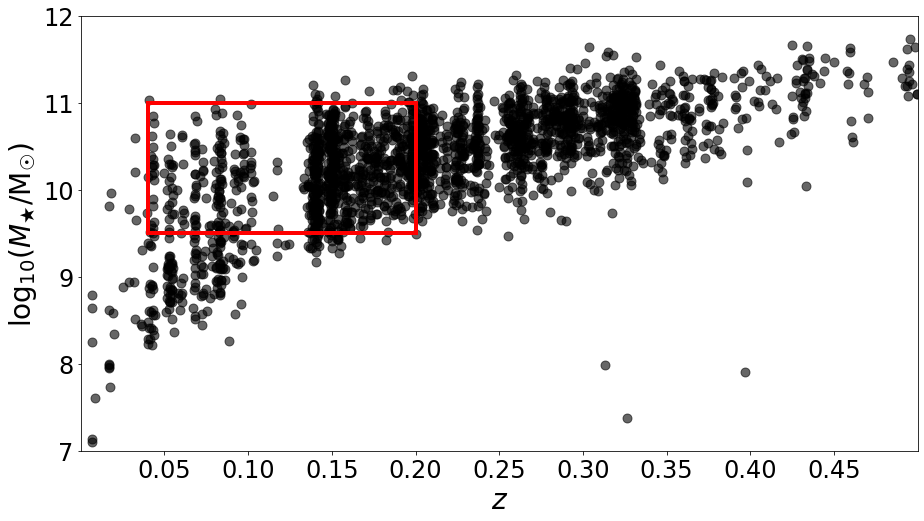}
    \caption{Distribution of redshift and stellar masses of the 2325 parent GAMA galaxies in the HSC-SSP Ultradeep Region. The Red rectangle shows our final volume-limited sample of 852 galaxies.}
    \label{fig:redshift_mass}
\end{figure}

To ensure that we can observe tidal features in our galaxies we limit our sample to the redshift range 0.04 $\leq$ \textit{z} $\leq$ 0.2, so that galaxies are close enough to be visually classified and the effects of evolution do not affect our analysis. We verified that the fraction of tidal features presented in Section \ref{sec:results} are unchanged as a function of redshift. To ensure that our sample of galaxies is consistent in their stellar masses over this redshift range, we apply stellar mass limits of 9.50~$\leq$~log$_{10}$($M_{\star}$/M$_{\odot}$)~$\leq$~11.00. Applying these limits, we get the final volume-limited sample of 852 galaxies shown in Figure \ref{fig:redshift_mass}. There is a visible gap in the redshift range 0.10 $\leq$ \textit{z} $\leq$ 0.14 with an absence of galaxies. This gap is caused by sky-line contamination in GAMA spectra at those wavelengths but does not affect the analysis presented here.
To ensure high completeness in our sample, we only include close pairs which have a companion galaxy that also meets our selection criteria. This ensures that our sample remains complete in both the low mass and high redshift regimes. After doing this, we are left with 46 galaxy close pairs whose members are both in our volume-limited sample. This is the sample presented through the remainder of this paper.

Due to the higher prevalence of low-mass galaxies (e.g. \citealt{Taylor2011GAMAMassEst,Baldry2012GAMAStellarMass}), it is more likely that higher-mass galaxies in our volume-limited sample could have a companion galaxy with mass ratio < 1:3, than the lower-mass galaxies, where such lower mass companion galaxies are below GAMA's detection limit of $r~\sim~19.8$~mag. To confirm that this effect does not bias our conclusions, we verified that all our conclusions remain qualitatively unchanged for a confirmation sample with stellar mass limits $10.00~\leq~$~log$_{10}(M_{\star}/\rm{M}_{\odot})~\leq~11.0$. For this confirmation sample, companion galaxies were not restricted to only being in the volume-limited sample, but instead were restricted to a stellar mass ratio $\geq$ 1:3 (major mergers) such that galaxies across our entire mass range had an equal probability of detecting a major close companion. This confirmation sample contains 605 total galaxies and 28 close pairs. The results presented in Section \ref{sec:results} are calculated using the larger sample of 852 galaxies as the larger range of stellar mass and colour in this sample allows for a more detailed analysis of mass and colour dependence in the tidal feature and close pair subsamples. All conclusions based on the results presented in Section \ref{sec:results} hold true for the confirmation sample unless otherwise discussed.

\subsection{Detection and Classification}
\label{sec:detect_class}
\begin{figure}
     \centering
     \begin{subfigure}[t]{0.42\columnwidth}
         \centering
         \includegraphics[width=\columnwidth]{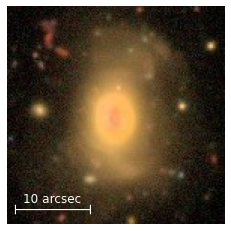}
     \end{subfigure}
     \hspace{0.5em}%
     \begin{subfigure}[t]{0.42\columnwidth}
         \centering
         \includegraphics[width=\columnwidth]{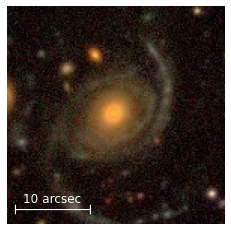}
     \end{subfigure}
     \vspace{0.5em}
     \begin{subfigure}[b]{0.42\columnwidth}
         \centering
         \includegraphics[width=\columnwidth]{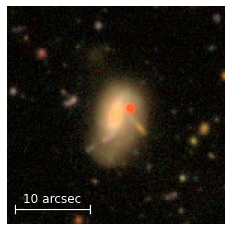}
     \end{subfigure}
     \hspace{0.5em}
     \begin{subfigure}[b]{0.42\columnwidth}
         \centering
         \includegraphics[width=\columnwidth]{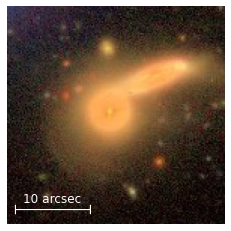}
     \end{subfigure}
        \caption{Example galaxies with tidal features in our sample (\textit{gri}-band cutout images). Top left: shells. Top Right: stream. Bottom left: asymmetric halo. Bottom Right: double nucleus}
        \label{fig:examples}
\end{figure}

The visual identification of the tidal features present in our galaxy sample will be based on a simplified form of the classification schemes used in \citet{Bilek2020MATLASTidalFeat} and \citet{Martin2022TidalFeatMockIm}. Their categories include:
\begin{enumerate}
    \item Shells: Concentric radial arcs or ring-like structures around a galaxy
    \item Tidal tails: Prominent, elongated structures expelled from the host galaxy. These usually have similar colours to that of the host galaxy.
    \item Stellar streams: Narrow filaments orbiting a host galaxy.
    \item Plumes or asymmetric stellar haloes: Diffuse features in the outskirts of the host galaxy, lacking well-defined structure like stellar streams or tails.
    \item Double nuclei: Galaxies which are visibly merging but where both objects are still clearly separated.
\end{enumerate}
The categories are not mutually exclusive, and galaxies with multiple features can belong to multiple categories simultaneously. Tidal tails and stellar streams share very similar morphologies and upon analysis of our sample appear to be indistinguishable. Because of this, we do not try to distinguish between these two types of tidal features in our analysis and label both types as stellar streams. Shells have been shown to primarily originate from minor mergers \citep{KadoFong2018HSCTidalFeat}, while stream-like features can originate from minor or major mergers (e.g. \citealt{Knierman2012Tails, Knierman2013Tails, Hendel2015TidalDebOrbit, Rodruck2016Tails}). We therefore expect that the use of tidal features will enable us to detect both major and minor mergers. Figure \ref{fig:examples} shows examples of each type of feature from our dataset. The classification was performed by the first author.

\newpage
\subsection{Defining a Colour Cut}
\label{sec:colour_cut}

In our analysis, we make use of \textit{g – i} colour as an indicator of morphology by using the tendency for redder galaxies to be quiescent early-types and bluer galaxies to be star-forming late-types (e.g. \citealt{Taylor2015GAMARedBlue}). Hence, by comparing the colours of the galaxies present in our samples, we can determine whether tidal feature or close pair samples are more likely to be associated with a particular galaxy colour.

\begin{figure}
    \centering
	\includegraphics[width=0.95\columnwidth]{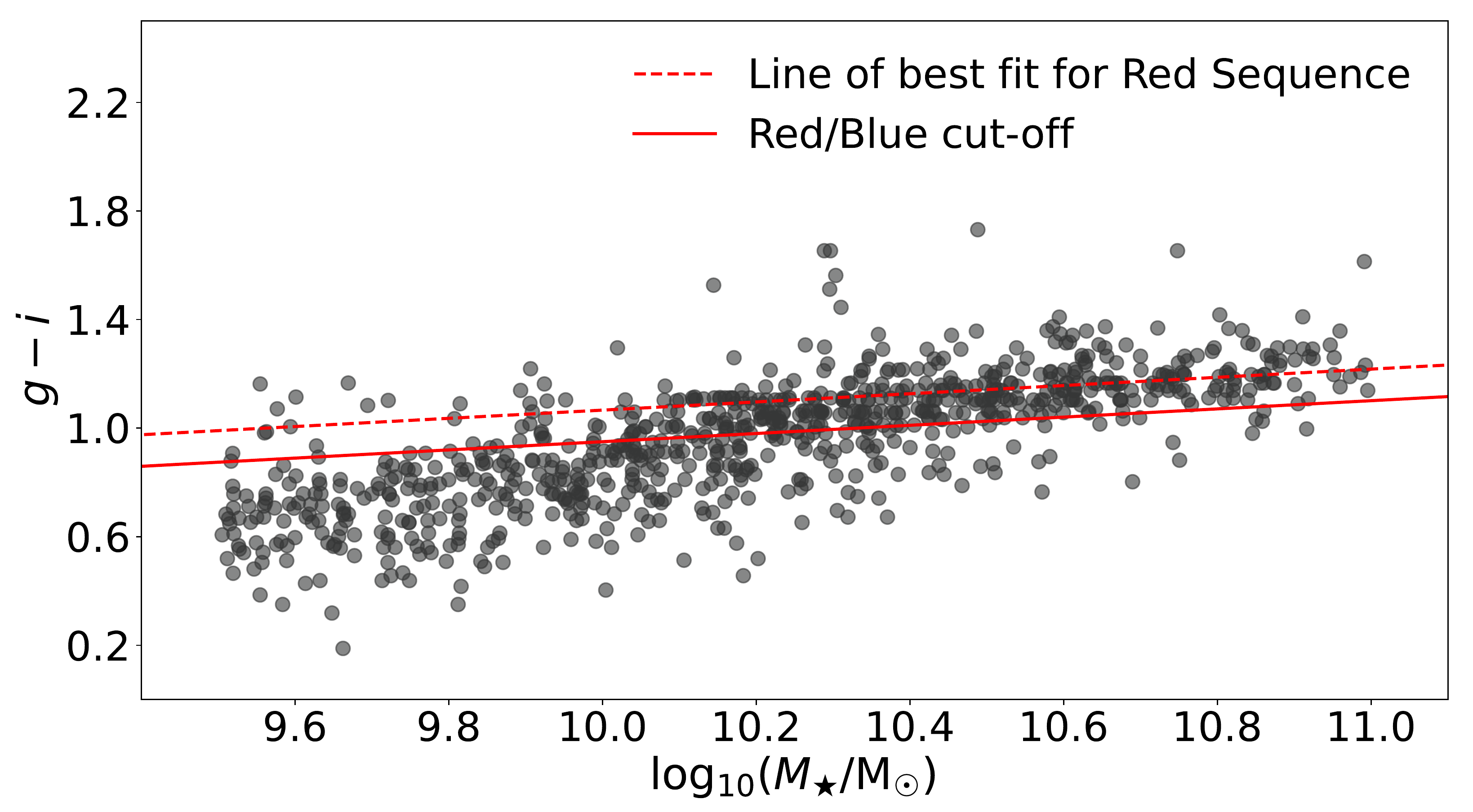}
    \caption{\textit{g – i} colour as a function of stellar mass. The dashed line shows the line of best fit for the Red Sequence and the solid line shows the separation between blue (below), and red (above) galaxies.}
    \label{fig:col_mass}
\end{figure}

 Of our 852 galaxies, six of them did not have magnitude information due to incomplete data in the HSC-SSP database so we did not include them in our colour cut calculations. These galaxies do not belong to any one particular subsample presented in Section \ref{sec:results}. To define a colour cut, we plotted the \textit{g – i} colour with respect to stellar mass for the 846 galaxies with colour information in our sample (Figure \ref{fig:col_mass}). Within this distribution the red sequence of galaxies is visible redwards of \textit{g – i}~$\sim$~1.0. We fit a straight line to the region 1.0~$\leq$~\textit{g – i}~$\leq$~1.4 to obtain the red sequence distribution and took the 1$\sigma$ root-mean-square scatter below the fitted line to be our red vs. blue galaxy cut-off point (e.g.  \citealt{Brough2017SAMIMass}). This red vs. blue cut-off is illustrated in Figure \ref{fig:col_mass} and follows the relationship:

\begin{equation}
          g-i = 0.15\times \textrm{log}_{10} (M_{\star}/\rm{M_{\odot}}) - 0.56
	\label{eq:colour_eq}
\end{equation}

We verify that the results presented in Section \ref{sec:results} hold true when the fit to the red sequence is varied. We explored three variations of this fit, firstly moving the lower limit of the \textit{g – i} fit to 1.05~$\leq$~\textit{g~–~i}~$\leq$~1.4, secondly by implementing a lower stellar mass limit log$_{10}(M_{\star}/\rm{M}_{\odot})~\leq~10.1$ to our fit, and thirdly by implementing both the lower \textit{g – i} limit and the lower stellar mass limit. The results presented in Section \ref{sec:results} remain qualitatively unchanged unless otherwise specified.
After applying our red vs. blue galaxy selection we find 453 red galaxies and 393 blue galaxies in our sample of 846 galaxies with colour information. Red galaxies are slightly more prevalent with a fraction of $f_{\rm{red}}$~=~0.54~$\pm$~0.02 compared to blue galaxies which have a fraction $f_{\rm{blue}}$~=~0.46~$\pm$~0.02. For our confirmation sample, red galaxies are more prevalent with a fraction of $f_{\rm{red}}$~=~0.70~$\pm$~0.02 compared to blue galaxies which have a fraction $f_{\rm{blue}}$~=~0.30~$\pm$~0.02. This is a direct consequence of the mass-colour relation and is an expected result of raising our lower stellar mass limit.

\subsection{Uncertainties}
\label{sec:unertainties}

Throughout this paper, the uncertainties on the presented fractional quantities are estimated using the beta distribution quantile technique to calculate confidence intervals \citep{Cameron2011ConfInt}. We use this method to calculate our confidence intervals as it is well-suited to smaller fractions and performs well for small-to-intermediate sample sizes.
All uncertainties in this paper will be given at a confidence level of \textit{c} = 0.683 equivalent to 1$\sigma$.

\section{Results}
\label{sec:results}

In this work, we wish to analyse how the selection of merging galaxies biases our understanding of galaxy interactions. Earlier studies focusing on galaxy interactions (e.g. \citealt{DePropris2005MillenniumPairs, Holwerda2011HIMorph, Robotham2014GAMAClosePair,KadoFong2018HSCTidalFeat}) have relied either on a spectroscopic close-pair sample, or on a visually-identified sample of interacting galaxies exhibiting tidal features, to draw their conclusions. Here we will compare the properties of a visually-identified sample of interacting galaxies exhibiting tidal features, and a spectroscopic close-pair sample drawn from the same parent sample. To do this we analyse the differences in the distribution of stellar mass, type of tidal feature, and colour of galaxies between the two samples as well as examine the properties of galaxies that appear in both samples. All quantities calculated here are summarised in Table \ref{tab:apend}.

\begin{table*}
\caption{Number, fraction of total, and mean stellar mass of galaxies in each sample described in Section \ref{sec:results} and the number, fraction, and mean colour of red and blue galaxies in each sample. The TFO, CPO, and CP+TF samples are the Tidal Feature Only, Close Pair Only, and Close Pair and Tidal Feature samples, respectively.}
\label{tab:apend}
\small
\centering
\begin{tabular}{p{2.4cm}>
{\centering}m{0.6cm}>
{\centering}m{1.5cm}>
{\centering}m{1.5cm}>
{\centering}m{0.6cm}>
{\centering}m{1.5cm}>
{\centering}m{1.5cm}>
{\centering}m{0.6cm}>
{\centering}m{1.5cm}>
{\centering\arraybackslash}m{1.5cm}}
\toprule
$~$ & {$N$} & {$f$} & {Mean log$_{10}$($M_{\star}$/M$_{\odot}$)} & {$N_{\rm{red}}$} & {$f_{\rm{red}}$} & {$\bar{x}_{\rm{red}}$\textit{(g-i)}} & {$N_{\rm{blue}}$} & {$f_{\rm{blue}}$} & {$\bar{x}_{\rm{blue}}$\textit{(g-i)}} \\
\midrule
Whole Sample & 852 & -- & 10.21 $\pm$ 0.01 & 453 & 0.54 $\pm$ 0.02 & 1.14 $\pm$ 0.01 & 393 & 0.46 $\pm$ 0.02 & 0.77 $\pm$ 0.01 \\
\\
Tidal Feature Sample & 198 & 0.23 $\pm$ 0.02 & 10.37 $\pm$ 0.03 & 121 & 0.62 $\pm$ 0.04 & 1.17 $\pm$ 0.01 & 74 & 0.38 $\pm$ 0.04 & 0.78 $\pm$ 0.02 \\
\\
Close Pair Sample & 80 & 0.09 $\pm$ 0.01 & 10.26 $\pm$ 0.04 & 49 & 0.62 $\pm$ 0.06 & 1.16 $\pm$ 0.02 & 30 & 0.38 $\pm$ 0.06 & 0.78 $\pm$ 0.03 \\
\\
TFO Sample & 156 & 0.18 $\pm$ 0.01 & 10.39 $\pm$ 0.03 & 91 & 0.59 $\pm$ 0.04 & 1.18 $\pm$ 0.01 & 62 & 0.41 $\pm$ 0.04 & 0.78 $\pm$ 0.02 \\
\\
CPO Sample & 38 & 0.04 $\pm$ 0.01 & 10.19 $\pm$ 0.05 & 19 & 0.51 $\pm$ 0.08 & 1.17 $\pm$ 0.04 & 18 & 0.49 $\pm$ 0.08 & 0.80 $\pm$ 0.03 \\
\\
CP+TF Sample & 42 & 0.05 $\pm$ 0.01 & 10.34 $\pm$ 0.05 & 30 & 0.71 $\pm$ 0.08 & 1.15 $\pm$ 0.02 & 12 & 0.29 $\pm$ 0.08 & 0.75 $\pm$ 0.05 \\
\\
Neither Sample & 616 & 0.72 $\pm$ 0.02 & 10.16 $\pm$ 0.01 & 313 & 0.51 $\pm$ 0.02 & 1.12 $\pm$ 0.01 & 301 & 0.49 $\pm$ 0.02 & 0.76 $\pm$ 0.01 \\
\\
\bottomrule
\end{tabular}
\end{table*}

\subsection{Tidal Feature and Close Pair Fractions}
\label{sec:tfcp_frac}
We calculate the fractions of tidal features and close pairs present in our sample of galaxies. The tidal feature fraction is defined as:

\begin{equation}
    f_{\rm{tidal}} = \frac{N_{\rm{tidal}}}{N_{\rm{gal}}}
\end{equation}
where $N_{\rm{gal}}$ is the total number of galaxies in our sample, and $N_{\rm{tidal}}$ is the total number of galaxies visually-classified as possessing tidal features in our volume-limited sample. We identify 198 galaxies with tidal features. This yields a tidal feature fraction of $f_{\rm{tidal}}$~=~0.23~±~0.02.
The close pair fraction is the fraction of our whole sample of galaxies that are present in the GAMA spectroscopic close pair sample. When using the close pair fraction to calculate galaxy merger rates, several correction factors are usually applied (e.g. \citealt{Patton2000ClosePairAccRates,DePropris2005MillenniumPairs,Holwerda2011HIMorph}). Given that the focus of this paper is to determine whether any observed galaxy possesses companion galaxies, not to calculate the galaxy merger rate or frequency of mergers, we use the uncorrected fraction of galaxies in close pairs, defined as:

\begin{equation}
    f_{\rm{pair}} = \frac{N_{\rm{pair}}}{N_{\rm{gal}}}
\end{equation}
where $N_{\rm{pair}}$ is the total number of galaxies in the GAMA close pair sample in our volume-limited sample. We identify 80 galaxies that are in close pairs where both galaxies are in our volume limit. This yields a close pair fraction of $f_{\rm{pair}}$~=~0.09~$\pm$~0.01.

To conduct our comparison of the tidal feature and close pair samples we wish to look at the galaxies that are in both samples or in one but not the other. To do this we split our tidal feature and close pair samples into three subsets: tidal feature only (TFO), close pair only (CPO), and close pair and tidal feature (CP+TF). The fourth subset contains galaxies which are not in close pairs nor possess tidal features, referred to as the Neither sample from here onward.

We calculate the fraction of galaxies in each subset. In our volume-limited sample of 852 galaxies, we identify 38 galaxies that are in close pairs only, 156 galaxies that only possess tidal features, 42 galaxies that are both in close pairs and possess tidal features, and 616 galaxies neither in close pairs nor with tidal features. The TFO, CPO, CP+TF, and Neither fractions are as follows

\begin{align}
    f_{\rm{TFO}}=0.18 \pm 0.01 \nonumber \\
    f_{\rm{CPO}}=0.04 \pm 0.01 \\
    f_{\rm{CP+TF}}=0.05 \pm 0.01 \nonumber \\
    f_{\rm{Neither}}=0.72 \pm 0.02 \nonumber
\end{align}

The fraction of galaxies in close pairs only, $f_{\rm{CPO}}$, and the fraction of galaxies both in close pairs and possessing tidal features, $f_{\rm{CP+TF}}$, are similar but these two subsets consist of different galaxies. The fraction of galaxies possessing tidal features, $f_{\rm{TFO}}$, is significantly larger than both $f_{\rm{CPO}}$ and $f_{\rm{CP+TF}}$, indicating that galaxies are significantly more likely to possess tidal features than be in close pairs, and 53 $\pm$ 6\% of galaxies in close pairs will also possess tidal features (see Table \ref{tab:apend}). We calculate the mean radial separations of galaxies in the CPO and CP+TF samples and find $\bar{r}_{sep}~$=~34.3~$\pm$~1.8~$h^{-1}$~kpc for the CPO sample, and $\bar{r}_{sep}~$=~21.7~$\pm$~1.7~$h^{-1}$~kpc for the CP+TF sample. Galaxies which are both in close pairs and possess tidal features therefore have significantly lower mean radial separations than galaxies which are only in close pairs.

In the following Sections we compare the galaxy properties for these four subsets to establish the properties of galaxies in close pairs and galaxies with tidal features. 

\subsection{Dependence on Stellar Mass}
\label{sec:dep_mass2}
In this Section we investigate whether galaxy merger samples detected through close pairs or the presence of tidal features depend on the stellar mass of the galaxies.

\begin{figure}
    \centering
	\includegraphics[width=0.85\columnwidth]{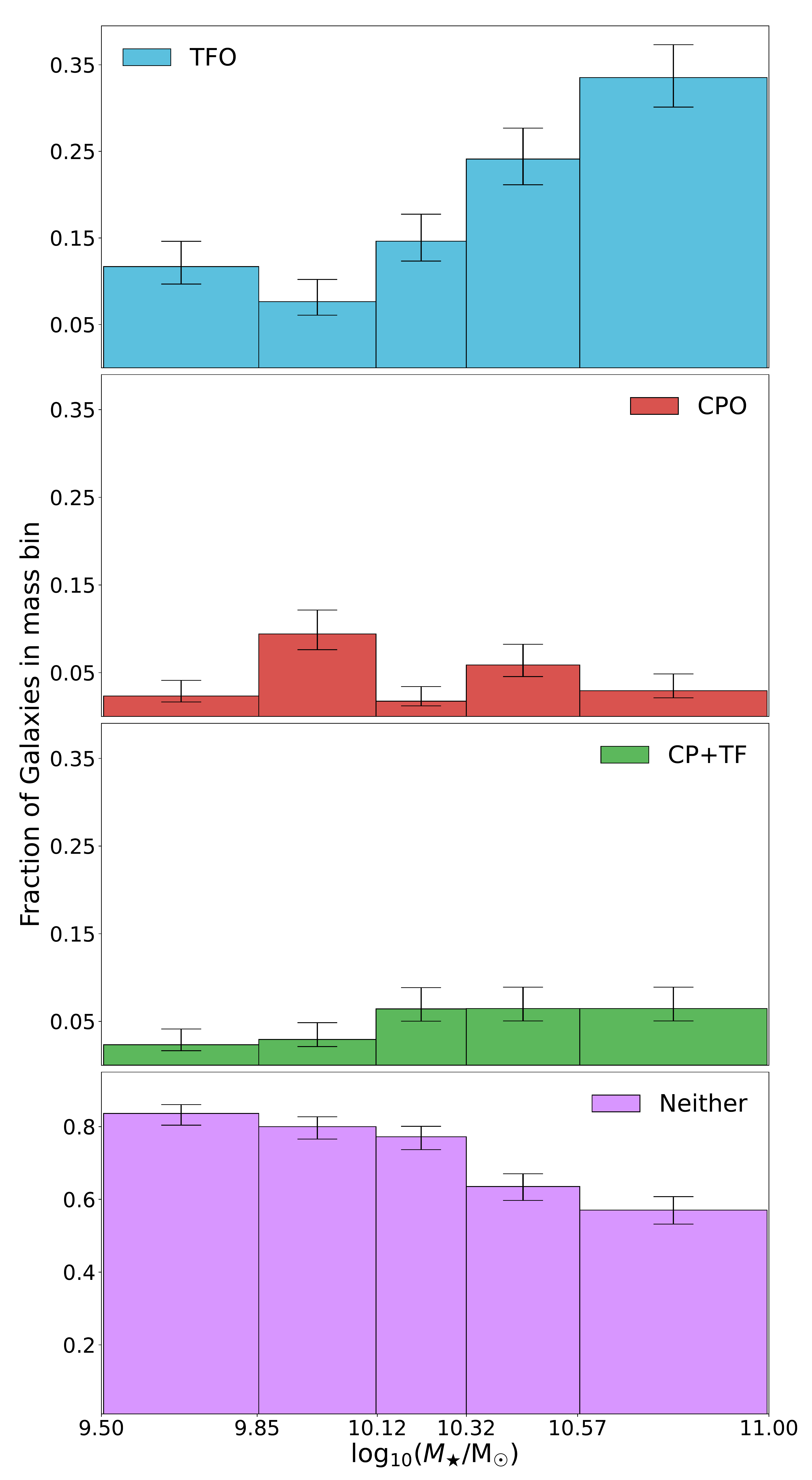}
    \caption{Fraction of galaxies as a function of stellar mass for the tidal feature only (TFO; top panel), close pair only (CPO; middle-top panel), close pair and tidal feature (CP+TF; middle-bottom panel), and Neither (bottom panel) subsamples. Each bin contains $\sim$170 parent galaxies. The TFO fraction shows a continuous increase with increasing stellar mass, whilst the Neither sample decreases.}
    \label{fig:bins3}
\end{figure}

Figure \ref{fig:bins3} illustrates the fractions of galaxies in the TFO, CPO, CP+TF, and Neither samples as a function of stellar mass. The fraction of TFO galaxies shows a continuous increase with stellar mass, increasing from $0.12~\pm~0.03$ for 9.5~$\leq$~log$_{10}$($M_{\star}$/M$_{\odot}$)~$\leq$~9.85 to $0.34~\pm~0.04$ for 10.57~$\leq$~log$_{10}$($M_{\star}$/M$_{\odot}$)~$\leq$~11. The fraction of galaxies in the CPO and CP+TF samples do not appear to have a significant relationship to the stellar mass within these galaxies, although galaxies in the CP+TF sample do have significantly higher mean stellar mass than galaxies in the Neither or CPO samples (Table \ref{tab:apend}). This difference in mean stellar mass for the various subsamples does not hold true for our confirmation sample as a direct consequence of the limited stellar mass range of this sample. The fraction of galaxies neither in close pairs nor with tidal features shows an inverse relationship with stellar mass, which is likely the inverse of the relationship between stellar mass and the fraction of galaxies with tidal features.

\subsection{Dependence on Galaxy Colour}
\label{sec:dep_col2}
\begin{figure}
    \centering
	\includegraphics[width=0.85\columnwidth]{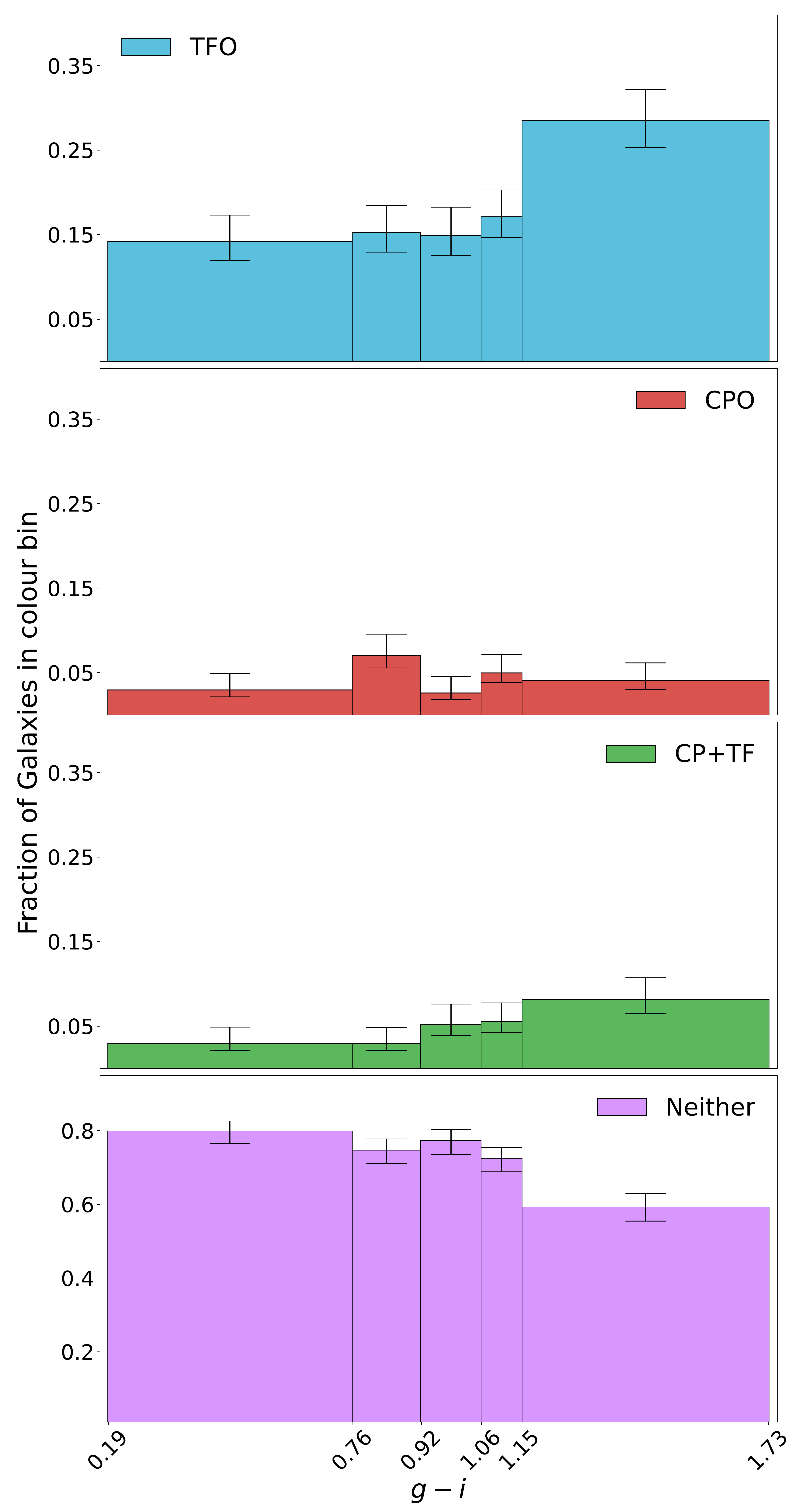}
    \caption{Fraction of galaxies as a function of \textit{g – i} colour for the TFO (top panel), CPO (middle-top panel), CP+TF (middle-bottom panel), and Neither (bottom panel) subsamples. Each bin contains $\sim$ 170 parent galaxies. Both the TFO and CP+TF sample detect redder galaxies than the CPO or Neither samples.}
    \label{fig:bins4}
\end{figure}

In this Section we investigate whether the \textit{g – i} colour of the galaxies is dependent on which sample they belong to: TFO, CPO, CP+TF, or Neither. We calculated the mean blue and red colour for the galaxies in each of our four subsamples, provided in Table \ref{tab:apend}. The CP+TF sample has a significantly higher number of red galaxies than blue galaxies, and this is also the case for the TFO sample, albeit less strong. When changing the conditions of our colour cut, as described in Section \ref{sec:colour_cut}, the CP+TF sample still has a significantly higher number of red galaxies, however the TFO sample is more dependent on the precise separation between red and blue galaxies. The fractions of red and blue galaxies in the TFO sample as approximately equal. To further our colour analysis we show in Figure \ref{fig:bins4} the fraction of galaxies in the TFO, CPO, CP+TF, and Neither samples in \textit{g – i} colour bins. Figure \ref{fig:bins4} confirms that both the TFO and CP+TF samples detect redder galaxies than the CPO or Neither samples. The galaxies in the Neither sample, on the other hand, are bluer than those in the CPO sample. This is consistent with the finding in Section \ref{sec:dep_mass2} that the CP+TF sample isolates the most massive, and hence reddest, galaxies from the galaxies in close pairs.

We find that the fractions of red and blue galaxies in samples of interacting galaxies (Table \ref{tab:apend}) are independent of whether the sample is a close pair or a tidal feature sample, with both samples detecting a higher fraction of red galaxies than blue galaxies. This is likely a consequence of galaxies with tidal features and galaxies in close pairs having higher average stellar masses than the non-interacting sample. This is confirmed by the absence of a difference in the fraction of red and blue galaxies between the tidal feature, close pair, and Neither sample in our confirmation sample.

\subsection{Dependence on Feature Type}
\label{sec:dep_feat_type}
In this Section we investigate whether the type of tidal feature exhibited by a galaxy is dependent on the \textit{g – i} colour or stellar mass of that galaxy. We also investigate whether there is a correlation between the type of tidal feature a galaxy exhibits and whether it belongs to the TFO or CP+TF sample. 

\begin{table}
	\centering
	\caption{Fraction and mean stellar mass of host galaxies with shells, streams, plumes, and double nuclei in our sample. Uncertainties are given to 1$\sigma$.}
	\label{tab:feature}
	\begin{tabular}{lccc}
		\toprule
		\thead{Feature Type} & \thead{Fraction of \\ all galaxies} & \thead{Mean \\ log$_{10}$($M_{\star}$/M$_{\odot}$)} & \thead{Mean \\ $g-i$ colour} \\
		\midrule
		Shell & 0.02 $\pm$ 0.01 & 10.56 $\pm$ 0.06 & 1.03 $\pm$ 0.05 \\
		Stream & 0.11 $\pm$ 0.01 & 10.42 $\pm$ 0.04 & 1.05 $\pm$ 0.02 \\
		Plume & 0.10 $\pm$ 0.01 & 10.44 $\pm$ 0.04 & 1.03 $\pm$ 0.03 \\
		Double Nucleus & 0.10 $\pm$ 0.01 & 10.33 $\pm$ 0.04 & 1.02 $\pm$ 0.03 \\
		\bottomrule
	\end{tabular}
\end{table}

\begin{figure}
    \centering
	\includegraphics[width=0.95\columnwidth]{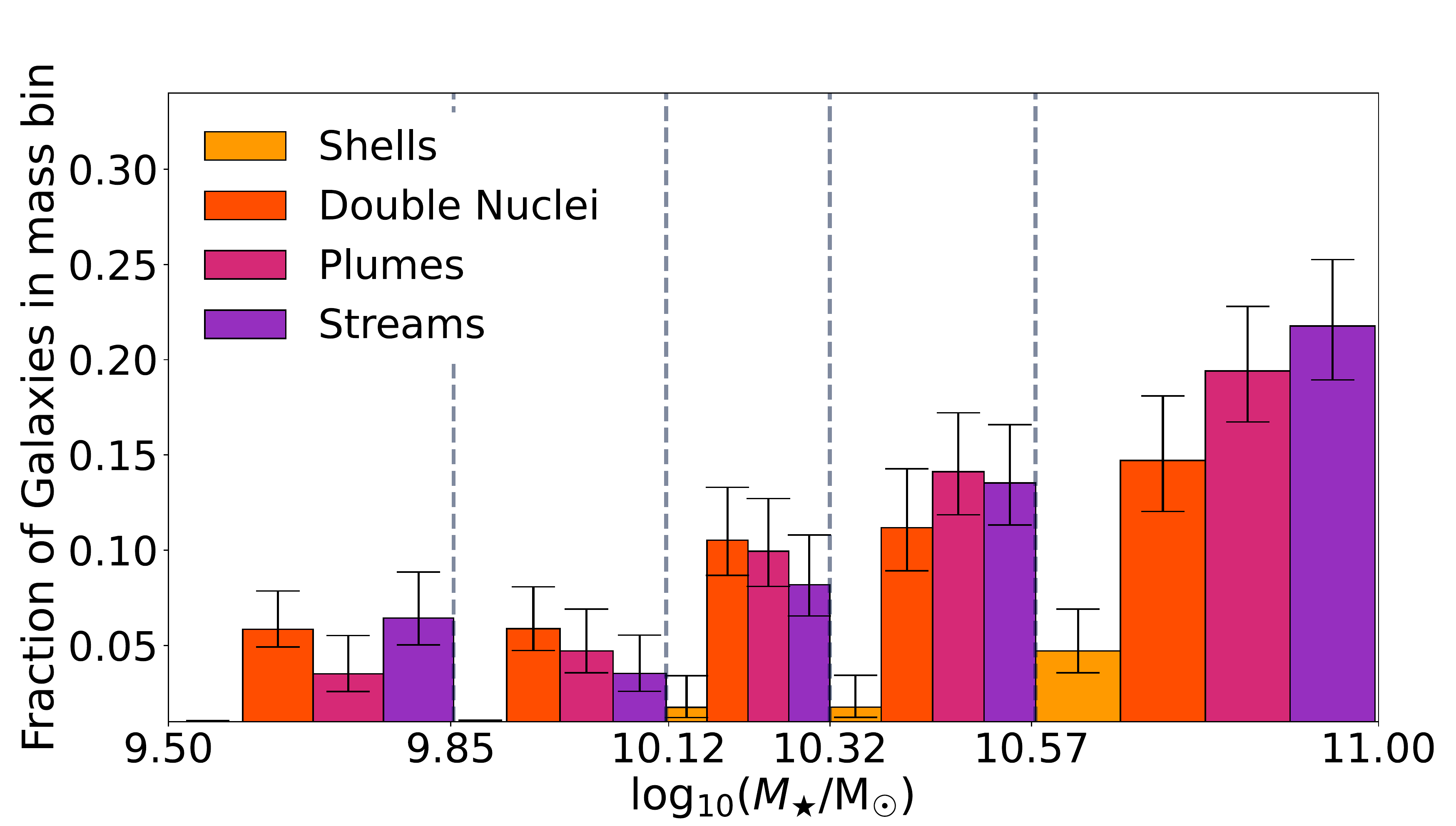}
    \caption{Fraction of galaxies hosting shell, double nucleus, plume, and stream features as a function of stellar mass. The dashed lines show the mass-bin edges, with each bin containing $\sim$ 170 parent galaxies (i.e. including those without tidal features). Shells are only present around higher mass galaxies whereas streams, plumes, and double nuclei are present around galaxies with a wide range of stellar masses.}
    \label{fig:bins5}
\end{figure}

\begin{figure}
    \centering
	\includegraphics[width=0.85\columnwidth]{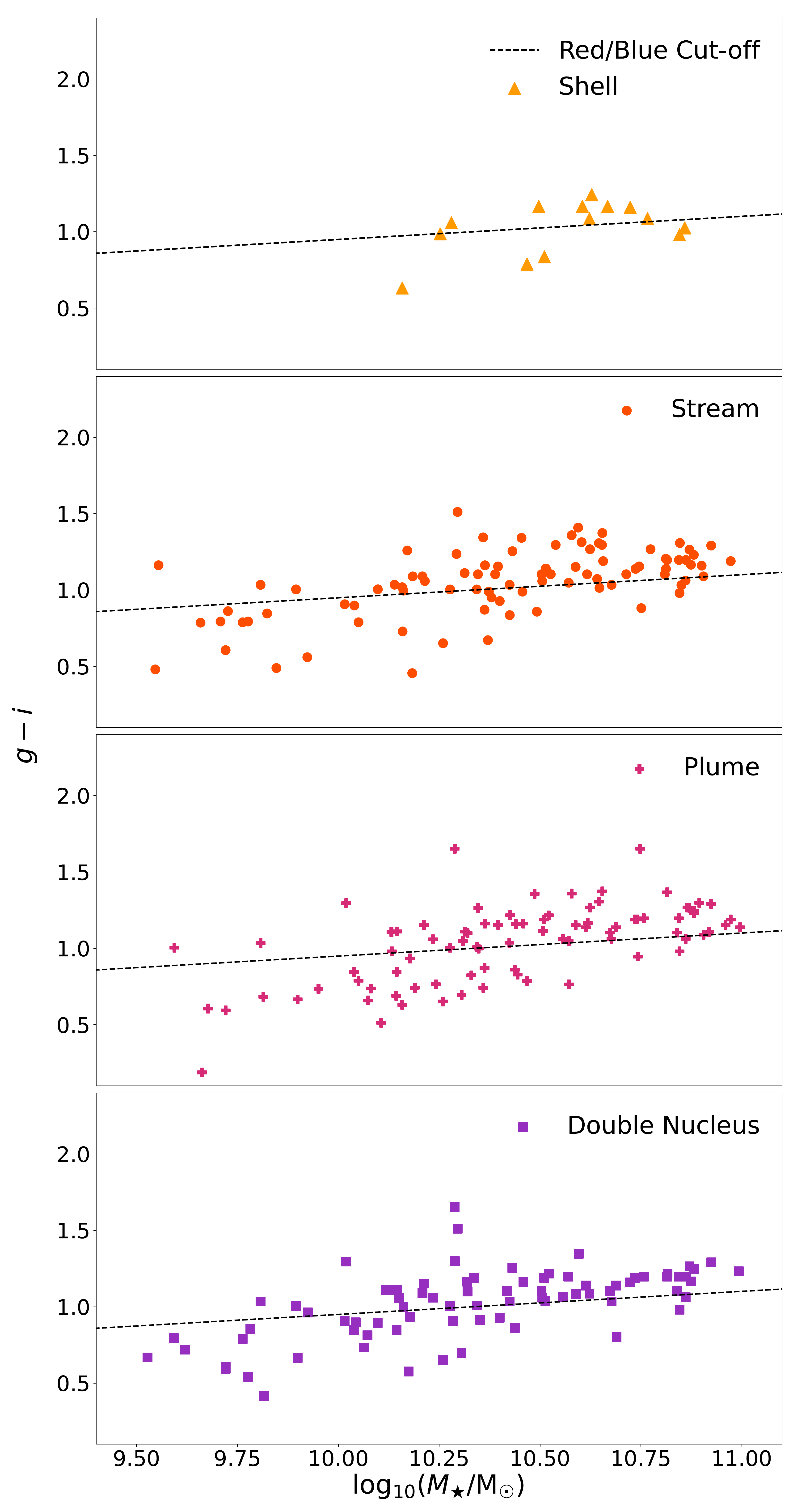}
    \caption{\textit{g – i} colour as a function of stellar mass for galaxies with shells (top), streams (middle-top), plumes (middle-bottom), and double nuclei (bottom). The dashed lines show our dividing line between red and blue galaxies. Shells are only present around higher mass galaxies close to the red vs. blue cut-off.}
    \label{fig:scatter3}
\end{figure}

Figure \ref{fig:bins5} illustrates the fractions of shells, streams, plumes, and double nuclei in our whole sample as a function of stellar mass. Figure \ref{fig:scatter3} shows the \textit{g – i} colour distribution as a function of galaxy stellar mass for galaxies with shells, streams, plumes, and double nuclei. Table \ref{tab:feature} gives the total fraction of galaxies with shells, streams, plumes, and double nuclei in our sample. These figures show that in our sample shells are only detected around higher mass galaxies where log$_{10}$($M_{\star}$/M$_{\odot}$) $\geq$ 10.10 and are usually found around galaxies close to the blue vs. red cut-off. Streams and plumes, on the other hand, share similar distributions, appearing around galaxies at a wide range of colour and mass, but, as seen in Figure \ref{fig:bins5}, increase in prevalence with increasing stellar mass. Double nuclei also appear in galaxies at a wide range of colour and mass and seem to show a weak increase in prevalence with increasing galaxy stellar mass.

\begin{figure}
     \centering
     \includegraphics[width=0.95\columnwidth]{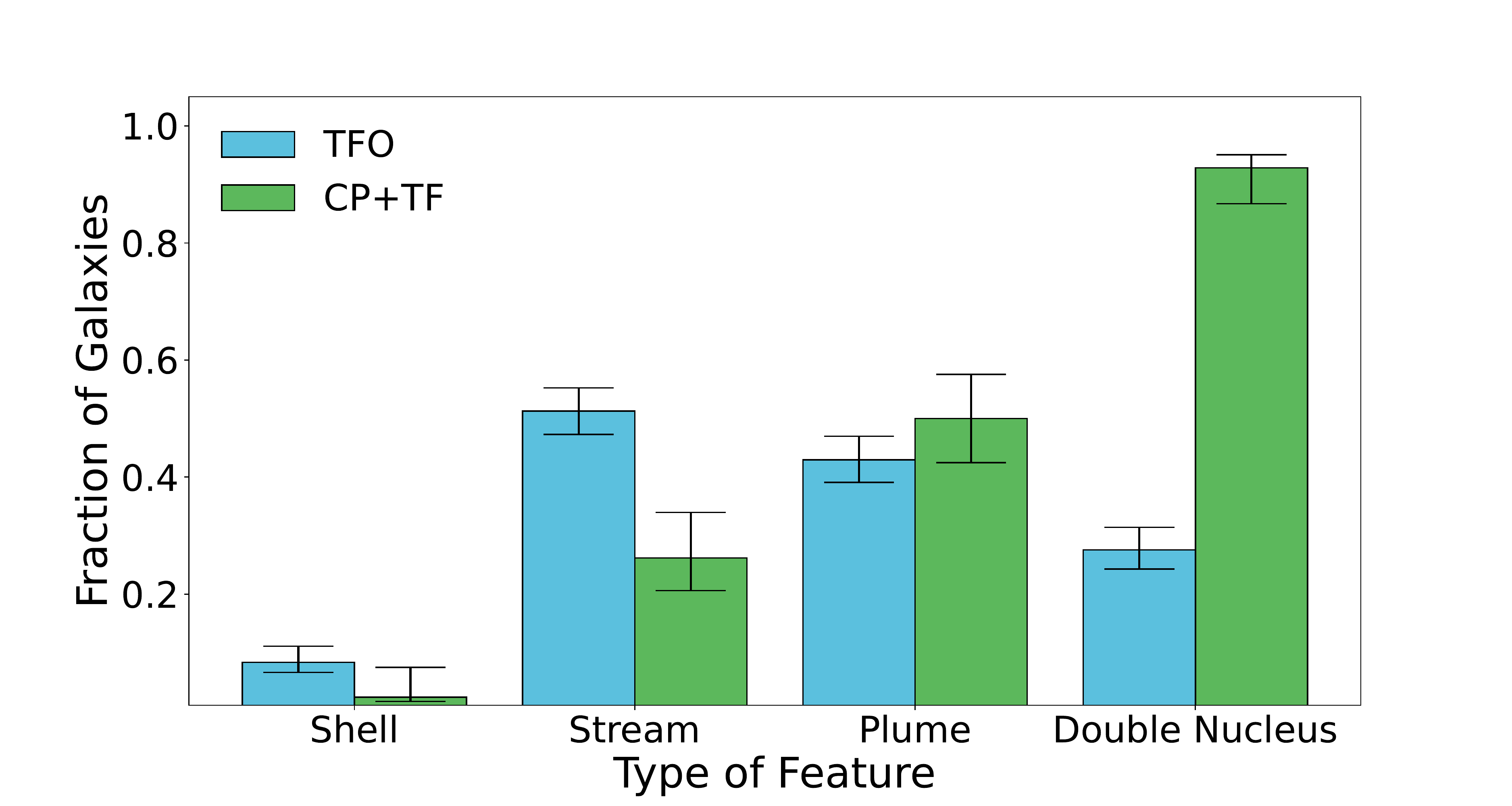}
     \caption{Fractions of the type of tidal feature for the TFO (green), and CP+TF (blue) samples. The CP+TF sample has a high fraction of galaxies with double nuclei. Streams, plumes and shells have similar fractions in the TFO and CP+TF sample.}
     \label{fig:bar3}
\end{figure}

Figure \ref{fig:bar3} compares the fractions of type of tidal feature in the TFO and CP+TF samples. In both samples, shells are the least frequent type of tidal feature, and plumes have similar prevalence across the two samples. Streams are more frequent in the TFO sample but the difference is not enough to be statistically significant, considering our 1$\sigma$ error bars. Galaxies with double nuclei are significantly more likely to also be classified as close pairs, with the fraction of galaxies in the TFO sample possessing double nuclei being 0.28~$\pm$~0.04 and rising to 0.93~$\pm$~0.06 for the CP+TF sample. Upon inspection of our images, all of the galaxies in the TFO sample with double nuclei appear to have a secondary galaxy that is either too close to the host galaxy or too small to be marked as a close pair by GAMA. This is likely due to the GAMA close pair sample being based on SDSS imaging which is shallower than HSC-SSP Ultradeep imaging with larger seeing ($\sim$2" vs. $\sim$1" full width at half-maximum; \citealt{Hill2011GAMASDSS, Aihara2019HSCSecondData}), and therefore small, faint, or close secondary galaxies would not be distinguishable from the host galaxy. Additionally, 2 of the 3 galaxies in the CP+TF samples which do not have a double nucleus do have a visible, but not obviously interacting, secondary galaxy in the cutout. The other galaxy in the CP+TF sample which was not visually classified as having a double nucleus has a larger radial separation than the area covered by our cutout \citep{Robotham2014GAMAClosePair}, meaning that the secondary galaxy is outside our field of view. From this, we conclude that whether a galaxy with tidal features is also in a close pair appears only to be strongly related to the double nucleus visual classification, rather than the presence of shell, plume, or stream features. 

\section{Discussion}
\label{sec:disc}
We have conducted an analysis of tidal features and close pairs for 852 GAMA galaxies located in the overlap of the GAMA G02 region and the HSC-SSP Ultradeep region in the redshift range 0.04~$\leq$~z~$\leq$~0.20 and mass range 9.50~$\leq$~log$_{10}$($M_{\star}$/M$_{\odot}$)~$\leq$~11.00. We calculated the tidal feature and close pair fractions and examined their dependence on galaxy stellar mass and colour, as well as quantifying the type and distribution of tidal features present in our sample. This Section compares these results with those obtained from previous analyses.

\subsection{Tidal Feature and Close Pair Fractions}
\label{sec:disc_cp_tf_frac}

We identify in our sample of 852 galaxies a total of 198 galaxies with visible tidal features and 80 galaxies in close pairs. This yields a total tidal feature fraction of $f_{\rm{tidal}}$~=~0.23~$\pm$~0.02 and a total close pair fraction of $f_{\rm{pair}}$~=~0.09~$\pm$~0.01. 

To compare the conclusions drawn from the close pair and tidal feature samples we also separated the tidal feature and close pair samples into three subsets: close pair only (CPO), tidal feature only (TFO), close pair and tidal feature (CP+TF), and a fourth subset containing galaxies which are not in close pairs nor possess tidal features, referred to as Neither. We identify 38 galaxies that are in close pairs only, 156 galaxies that only possess tidal features, 42 galaxies that are both in close pairs and possess tidal features, and 616 galaxies that are neither in close pairs nor possess tidal features. This yields the fractions $f_{\rm{CPO}}$ = 0.04 $\pm$ 0.01, $f_{\rm{TFO}}$ = 0.18 $\pm$ 0.01, $f_{\rm{CP+TF}}$ = 0.05 $\pm$ 0.01, and $f_{\rm{Neither}}$ = 0.72 $\pm$ 0.02.

\begin{figure}
    \centering
	\includegraphics[width=0.95\columnwidth]{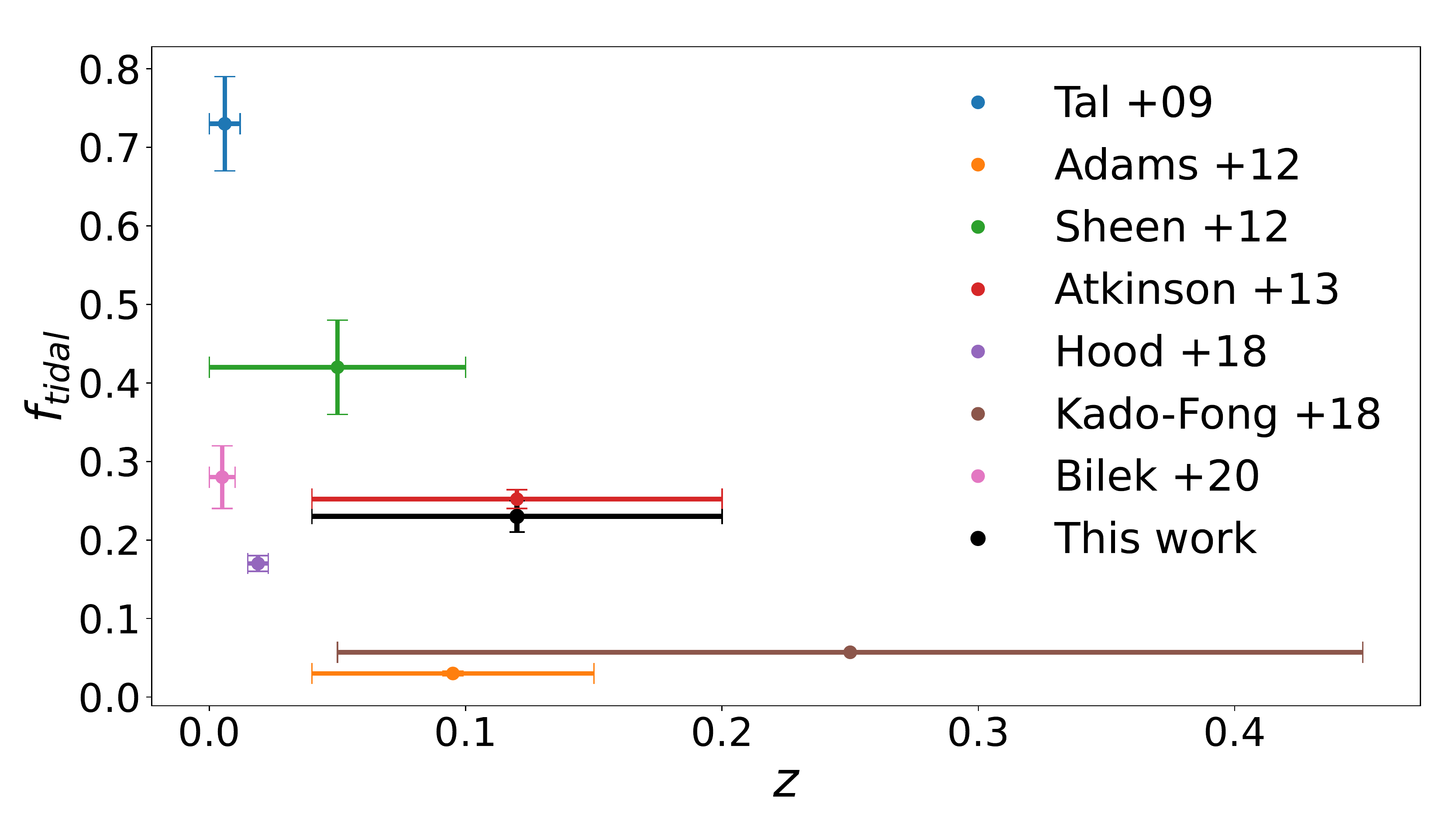}
    \caption{Tidal feature fractions in the literature. The horizontal error bars indicate the redshift range of the sample studied. Our results are in excellent agreement with those of \citet{Atkinson2013CFHTLSTidal} and \citet{Bilek2020MATLASTidalFeat}. The other fractions are calculated for samples with different redshift, stellar mass, and imaging depths limits to this work (see text).}
    \label{fig:past_lit}
\end{figure}

Previous tidal feature analyses found tidal feature fractions of $f_{\rm{tidal}}$~=~0.25~$\pm$~0.01 \citep{Atkinson2013CFHTLSTidal}, $f_{\rm{tidal}}$ = 0.28 $\pm$ 0.04 \citep{Bilek2020MATLASTidalFeat}, $f_{\rm{tidal}}$ = 0.17 $\pm$ 0.01 \citep{Hood2018RESOLVETidalFeat}, $f_{\rm{tidal}}$~=~0.73 \citep{Tal2009EllipGalTidalFeat}, $f_{\rm{tidal}}$ = 0.057 $\pm$ 0.002 \citep{KadoFong2018HSCTidalFeat}, $f_{\rm{tidal}}$~=~0.03~$\pm$~0.003 \citep{Adams2012EnvDep}, $f_{\rm{tidal}}$ = 0.42 $\pm$ 0.06 \citep{Sheen2012PostMergeSigs}. These results are compared in Figure \ref{fig:past_lit}. Tidal feature fractions show a strong dependence on the redshift, stellar mass, and imaging depths limits used in the analysis. The works which were carried out on samples with similar redshift, stellar mass, and imaging depth limits to this work are found to be in good agreement with our findings (e.g. \citealt{Atkinson2013CFHTLSTidal, Bilek2020MATLASTidalFeat}). Those which had shallower observing depths than the HSC-SSP Ultradeep layer used here, making faint tidal features much harder to detect, found lower tidal feature fractions (e.g. \citealt{Hood2018RESOLVETidalFeat, KadoFong2018HSCTidalFeat}). Moreover, \citet{KadoFong2018HSCTidalFeat} used a wider redshift range in their work, with an upper limit of \textit{z} = 0.45, finding the detection of tidal features increasingly difficult for lower mass galaxies at these high redshifts. The significantly higher tidal feature fraction of \citet{Tal2009EllipGalTidalFeat} can be attributed to the fact that their sample consisted only of massive, red, early-type galaxies, precisely the type of galaxies we found were most likely to possess tidal features. The galaxy samples studied by \citet{Adams2012EnvDep} and \citet{Sheen2012PostMergeSigs} were composed of massive, spheroidal galaxies located in clusters. Therefore, our results are not directly comparable due to these different sample selections. Despite studying similar samples of galaxies, the tidal feature fraction found in \citet{Adams2012EnvDep} and \citet{Sheen2012PostMergeSigs} are significantly different, likely due to the different surface brightness limits in these studies.

Although there are numerous works studying systems of merging galaxies through analysis of close pair samples, such as those of \citet{Darg2010GalZoo} and \citet{Robotham2014GAMAClosePair}, no value for the total close pair fraction could be found in those references. However, in their work analysing the stellar mass dependence of GAMA close pairs with masses 8~$\leq$~log$_{10}$($M_{\star}$/M$_{\odot}$)~$\leq$~12 in the redshift range 0.05~$\leq$~$z$~$\leq$~0.20,  \citet{Robotham2014GAMAClosePair} calculated the fraction of galaxies involved in major (mass ratio <  3 : 1) close pair interactions. They found an upper limit on the major close pair fraction of f$_{\rm{major}}$~=~0.09~–~0.13 in our mass range of interest of 9.50~$\leq$~log$_{10}$($M_{\star}$/M$_{\odot}$)~$\leq$~11.00. In our confirmation sample of 605 galaxies we found 43 galaxies involved in major mergers, giving a major close pair fraction of $f_{\rm{major}}$~=~0.07~$\pm$~0.01, in agreement within our 3$\sigma$ uncertainty. 

\citet{DePropris2007MillenniumClosePairAsym} studied the connection between close pairs and asymmetry using a sample of 3184 galaxies from the Millenium Galaxy Catalogue (MGC) and found a close pair fraction of $0.041~\pm~0.004$. This is lower than the close pair fraction calculated here however, their close pair selection criteria required that pairs had $r_{sep}~$<~20~$h^{-1}$~kpc, significantly smaller than our limit of $r_{sep}~$<~50~$h^{-1}$~kpc. \citet{Holwerda2011HIMorph} studied close pairs in the Westerbork observations of neutral Hydrogen in Irregular and SPiral galaxies (WHISP) sample. In the sample of 339 galaxies, they find 24 galaxies with close companions, resulting in a close pair fraction of $0.07~\pm~0.02$, but 13\% based on HI disturbance. WHISP is a targeted survey, based on the Uppsala General Catalogue. Although this close pair fraction is in agreement within 3$\sigma$ of the fraction calculated here, its lower value can be explained by the brighter limiting surface brightness of the Uppsala General Catalogue.


\subsection{Dependence on Stellar Mass}
\label{sec:disc_dep_mass}
We analysed the stellar mass dependence of our CPO, TFO, CP+TF, and Neither samples in Section \ref{sec:dep_mass2}. We found that the fraction of TFO galaxies shows a continuous increase with stellar mass. We also found that the fraction of galaxies in the CPO and CP+TF samples do not appear to have a significant relationship to the stellar mass within these galaxies, although galaxies in the CP+TF sample do have significantly higher mean stellar mass than those in the Neither sample.

In their study of CFHT Legacy Survey galaxies, \citet{Atkinson2013CFHTLSTidal} found a strong relationship between the stellar mass of galaxies and the occurrence of tidal features, finding that tidal features such as shells, stellar streams and tidal tails occur much more frequently in galaxies with stellar masses log$_{10}$($M_{\star}$/M$_{\odot}$) > 10.5. Similarly, \citet{Bilek2020MATLASTidalFeat} found a statistically significant increase of the incidence of shells, streams, and disturbed outer isophotes with the increasing stellar mass of the galaxy. They hypothesised that this increase in tidal feature occurrence at higher masses stems from the fact that more massive galaxies have stronger gravitational attraction and stronger tidal forces, leading to higher rates of tidal features. The stellar mass dependence of the total tidal feature fraction and the tidal feature fraction in our TFO sample that we observe are in good agreement with those findings. 

\citet{Darg2010GalZoo} studied 3003 close pairs of galaxies drawn from the SDSS in the redshift range 0.005~<~$z$~<~0.1 and stellar mass range 8.0~$\leq$~log$_{10}$($M_{\star}$/M$_{\odot}$)~$\leq$~12.0. They found that galaxies in close pairs are on average more massive than non-interacting galaxies. This is corroborated by our study, where galaxies in close pairs have an average mass log$_{10}$($M_{\star}$/M$_{\odot}$)~=~10.26~$\pm$~0.04 and non-interacting galaxies have a significantly lower average mass of log$_{10}$($M_{\star}$/M$_{\odot}$)~=~10.16~$\pm$~0.01.
In their study analysing the stellar mass dependence of close pairs, \citet{Robotham2014GAMAClosePair} find no evidence of a relationship between the stellar mass of galaxies and the likelihood of galaxies being in close pairs for the stellar mass range 9.50 $\leq$ log$_{10}$($M_{\star}$/M$_{\odot}$) $\leq$ 11.00. This is consistent with our findings regarding the absence of a relationship between galaxy stellar mass and the close pair fraction in both the whole close pair sample and the CPO and CP+TF samples.

\citet{Blumenthal2020IllustrisInteract} used the hydrodynamical cosmological simulation IllustrisTNG-100 \citep{Nelson2019IllustrisPDR} to test the use of visual classification of tidal features to detect interacting galaxy systems. They generated 446 galaxy pairs which were identified as interacting based on their trajectories (non-Visually Identified Pairs). The pairs were also visually classified as interacting or non-interacting based on the presence of tidal features (Visually Identified Pairs). They found that the Visually Identified Pairs tend to have higher stellar masses and have undergone a close passage twice as recently as the non-Visually Identified Pairs. This is consistent with our finding that the number of galaxies possessing tidal features increases continuously with increasing stellar mass.

\subsection{Dependence on Galaxy Colour}
\label{sec:disc_dep_col}
We investigated the galaxy colour dependence of our CPO, TFO, CP+TF, and Neither subsamples in Section \ref{sec:dep_col2}. We found that the TFO, and CP+TF subsamples possess a larger fraction of red, and therefore more likely to be early-type, galaxies, whilst the Neither subsample has an equal fraction of red and blue galaxies.

Both \citet{Atkinson2013CFHTLSTidal} and \citet{Bilek2020MATLASTidalFeat} found that all types of tidal features are more common in red galaxies than in blue galaxies, with a bias toward massive early-type galaxies. \citet{Lotz2011MajorMinorRate} found that three galaxy formation models, including two semi-analytical models and one abundance matching model, predict that at low redshifts, the population of merging galaxies is dominated by mergers with low gas fractions, associated with red, early-type galaxies as opposed to the high gas fractions associated with blue, late-type galaxies. Both the conclusions of visual studies, such as those of \citet{Atkinson2013CFHTLSTidal} and \citet{Bilek2020MATLASTidalFeat}, and the predictions from hydrodynamical simulations, such as those of \citet{Lotz2011MajorMinorRate} corroborate our findings that the merging populations in the local Universe are dominated by red early-type galaxies rather than blue late-type galaxies. However, the conclusions from these works and this paper contradict those of \citet{Hood2018RESOLVETidalFeat}. \citet{Hood2018RESOLVETidalFeat} found that tidal features are more likely to occur in gas-rich (19$\%$) galaxies, associated with late-type galaxies, than gas-poor (13$\%$) galaxies, associated with early-type galaxies. However, this discrepancy can be attributed to the \citet{Hood2018RESOLVETidalFeat} sample being focused on nearby ($z~\sim~0.02$) galaxies, allowing them to detect more low-mass, and hence gas-rich, galaxies. They also used the gas-to-stellar mass ratio for galaxies in their sample, leading to a different definition between gas content and colour for their sample than the one assumed in these analyses. 

\citet{Darg2010GalZoo} concluded that galaxy close pairs exhibit a higher spread in both red and blue colours compared to galaxies not in close pairs. They also found that below stellar masses of log$_{10}$($M_{\star}$/M$_{\odot}$)~$\sim$~10.5 elliptical galaxies were rare, regardless of whether these galaxies were in close pairs or not. However, the data used by \citet{Darg2010GalZoo} was based on much shallower observations (SDSS), making low-mass elliptical galaxies harder to detect. We find no evidence for a rarity of red galaxies below this stellar mass in our analysis (Figure \ref{fig:col_mass}). However, we are using galaxy colour as an indicator of morphology which is not a perfect identifier (e.g. \citealt{Taylor2015GAMARedBlue}). We also do not find evidence that galaxies in close pairs exhibit a higher spread in colour at either the red or blue ends compared to galaxies not in close pairs. We instead find that galaxies in our Neither sample appear to be slightly bluer than galaxies in close pairs.

\subsection{Dependence on Type of Feature}
\label{sec:disc_feat_type}
In Section \ref{sec:dep_feat_type} we calculated the prevalence of shells, streams, plumes, and double nuclei in our sample, finding fractions $f_{\rm{shell}}$~=~0.02~$\pm$~0.01, $f_{\rm{stream}}$ = 0.11 $\pm$ 0.01, $f_{\rm{plume}}$ = 0.10 $\pm$ 0.01, and $f_{\rm{double~nucleus}}$~=~0.10~$\pm$~0.01 for shells, streams, plumes, and double nuclei, respectively. We analysed the distribution of shells, streams, plumes, and double nuclei as a function of \textit{g – i} colour and galaxy stellar mass to establish whether a correlation exists between the type of tidal feature present and the properties of the host galaxies. We also compared the fractions of shells, streams, plumes, and double nuclei between our TFO and CP+TF samples to determine whether there is a dependence between the type of tidal feature exhibited and whether galaxies are also in close pairs. We found that shells were the least common type of tidal feature, only found around higher mass galaxies, log$_{10}$($M_{\star}$/M$_{\odot}$)~$\geq$~10.10. In contrast, streams and plumes appeared around galaxies with a wide range of colour and mass, but appeared to increase in prevalence with increasing stellar mass. Double nuclei also appeared in galaxies with a wide range of colour and mass and appeared to show a weak increase in prevalence with increasing galaxy stellar mass. We also found that galaxies with double nuclei were significantly more likely to also be classified as close pairs, with the fraction of galaxies in the TFO sample possessing double nuclei being 0.28~$\pm$~0.04 and rising to 0.93~$\pm$~0.06 for the CP+TF sample. Galaxies in the TFO sample with double nuclei appear to have secondary galaxies which were either too close to the host galaxy or too small to be marked as close pairs in the GAMA survey.

In their study of visually-classified HSC-SSP galaxies with tidal features, \citet{KadoFong2018HSCTidalFeat} found shells to be the least common type of tidal feature with an occurrence fraction of $f_{shell}$~=~0.01 and found that galaxies which host shells have predominantly redder \textit{g~–~i} colours and greater stellar masses than galaxies without tidal features. They also found a significant deficit of shells around late-type galaxies with the majority of shell hosts being early-type galaxies. They found a stream occurrence fraction of $f_{stream}$~=~0.05 and that galaxies which exhibit streams span the full range of stellar masses and colours present in the non-interacting sample at a given redshift. This fraction of galaxies possessing streams is lower than we find here but this can be attributed to the shallower observations of the HSC-SSP Wide layer used in their analysis. Our observations agree with those of \citet{KadoFong2018HSCTidalFeat} with respect to shells being the least common type of tidal feature and being found exclusively around galaxies with high stellar masses. However, the colours of the galaxies which host shells in our sample are usually close to the blue vs. red cut-off as opposed to being found predominantly around redder galaxies. This difference could be attributed to the shells having bluer colours than their host galaxies \citep{KadoFong2018HSCTidalFeat}, and the HSC-SSP Ultradeep photometry incorporating these bluer colours with the host galaxy, causing the galaxy to appear bluer.

In agreement with our results, \citet{Atkinson2013CFHTLSTidal} also found shells to be the least common type of tidal feature in their study of CFHT Legacy Survey galaxies. \citet{Bilek2020MATLASTidalFeat}, however, found a prevalence of 10-15$\%$ of shells, streams, and plumes in their study of tidal features. This can be attributed to their sample being composed only of massive early-type galaxies and thus being primarily high stellar mass, red galaxies which this work as well as \citet{Atkinson2013CFHTLSTidal} and \citet{KadoFong2018HSCTidalFeat} have found to be the most likely type of galaxy to host shells.

\citet{Martin2022TidalFeatMockIm} analysed the prevalence of different classes of tidal features as a function of stellar mass and limiting surface brightness, using mock images from the \texttt{NEWHORIZON} hydrodynamical simulation. They found that at brighter limiting surface brightnesses ($\mu_{r}<$~28~mag~arcsec$^{-2}$) streams and tails did increase in prevalence with increasing stellar mass but at fainter limiting surface brightnesses ($\mu_{r}\sim$~35~mag~arcsec$^{-2}$) showed similar prevalence across their entire stellar mass range ($\sim$~50$\%$). This suggests that the prevalence of streams and tails may not be dependent on galaxy stellar mass, but instead, since they are the faintest types of tidal features, are simply not visible below a certain limiting surface brightness. However, in agreement with our findings, \citet{Martin2022TidalFeatMockIm} found that shells did increase in prevalence with increasing stellar mass regardless of the limiting surface brightness.

\subsection{Merger Stages Detected by Close Pair and Tidal Feature Samples}
\label{sec:disc_merger_stages}
In this Section we examine the overlap (CP+TF samples) and divide (CPO and TFO samples) between our tidal feature and close pair samples to determine whether these two methods of detecting merging galaxy samples detect the same populations of merging galaxies. 
If close pair and tidal feature samples detected the same populations of merging galaxies, we would expect the overlap between our samples to be significantly greater than we find. However, we do find that galaxies in close pairs will also possess tidal features about 53$\%$ of the time ($N_{\rm{CP+TF}}/N_{\rm{All~close~pairs}}=49/93$; see Table \ref{tab:apend}) and that there is overlap between the galaxy populations detected by close pair and tidal feature samples. Additionally, the high fraction of double nuclei present in the CP+TF sample (0.93 $\pm$ 0.06) strongly suggests that this dual-identified sample detects close pairs with smaller separations than the CPO close pairs, where dynamical friction is already strong enough to have pulled out stellar material from the galaxies. This was confirmed by calculating the mean radial separation of galaxies in the CPO and CP+TF samples. We found $\bar{r}_{sep}~$=~34.3~$\pm$~1.8~$h^{-1}$~kpc for the CPO sample, and $\bar{r}_{sep}~$=~21.7~$\pm$~1.7~$h^{-1}$~kpc for the CP+TF sample, confirming that the CP+TF sample does indeed detect close pairs with significantly smaller separations than the CPO sample. This is in excellent agreement with the conclusions of \citet{DePropris2007MillenniumClosePairAsym} who studied 3184 galaxies from the Millenium Galaxy Catalogue \citep{Liske2003MillenniumLumFunc} and found an increase in galaxy asymmetry (e.g. tidal features) with decreasing pair separation.

\citet{Lotz2011MajorMinorRate} used hydrodynamical merger simulations to estimate the observability timescale (T$_{\rm{obs}}$) of ongoing mergers and found close pairs to be observable for $\sim$~600~Myr in the pre-merger stage. \citet{Holwerda2011HIMorph} argue mergers are visible for longer based on HI disturbance which is consistent with \citet{Huang2022HSCTidalFeatETG} and our observations here that mergers are visible for longer in tidal features. \citet{Lotz2011MajorMinorRate} also found that galaxies exhibiting morphological disturbances such as tidal features are often observed at a late stage in the merger process, where two nuclei are counted as a single object; while galaxies classified as close pairs are often observed at an early stage in the merger process. The findings of \citet{Robotham2014GAMAClosePair} corroborated this idea, with their analysis finding that highly-disturbed close pairs will, on average, be more likely to merge on shorter time-scales than close pairs with no signs of visual disturbance. This is because they find galaxies in highly-disturbed close pairs to typically have smaller separations than galaxies in non-disturbed close pairs.

Our results are in good agreement with the conclusions of \citet{Lotz2011MajorMinorRate} and \citet{Robotham2014GAMAClosePair}, with the CPO sample likely detecting early-stage mergers, where two separate galaxies are still visible, and the TFO sample detecting predominantly late-stage mergers, where generally only one galaxy nucleus remains visible. Both the close pair sample and tidal feature sample are more likely to detect red galaxies but the tidal feature sample is more likely to detect merging galaxies with slightly higher stellar masses, with mean stellar masses, log$_{10}$($M_{\star}$/M$_{\odot}$)~=~10.39~$\pm$~0.03 and 10.19~$\pm$~0.05, for the TFO and CPO samples respectively. The tendency for the tidal feature sample to detect galaxies with higher stellar masses is expected if the galaxy mergers detected are predominantly late-stage mergers which have already accreted stellar mass from their merging companion. The CP+TF sample appears to detect intermediate-stage mergers, with 93 $\pm$ 6$\%$ of galaxies in this sample possessing visible double nuclei, compared to only 28 $\pm$ 4$\%$ of galaxies in the TFO sample. Of the 3 galaxies in the CP+TF sample that do not possess visible double nuclei, one of them has a secondary galaxy too far away to detect in our postage stamp images (60~$\times$~60"; $\sim$100kpc), and the other two have a secondary galaxy that is not obviously interacting. We expect the galaxies in the CP+TF sample to be more likely to merge on shorter time-scales than galaxies in the CPO sample, as per the findings of \citet{Robotham2014GAMAClosePair}.

\section{Conclusions}
\label{sec:conc}
We have analysed a volume-limited sample of 852 galaxies with spectroscopic redshifts between 0.04~$\leq$~$z$~$\leq$~0.20 from the region of overlap between the G02 region of the GAMA survey and the Ultradeep layer of the HSC-SSP survey. In this paper we have compared the galaxy properties for a sample of merging galaxies detected by visual classification of tidal features and a sample of merging galaxies detected spectroscopically via close pairs. This is the first time that such deep images have been available for a sample of galaxies that also has a sample of close pairs defined, allowing us to undertake this analysis on a homogeneous sample of galaxies.

\begin{itemize}
    \item We identified 198 galaxies possessing tidal features, resulting in a tidal feature fraction of $f_{\rm{tidal}}$ = 0.23 $\pm$ 0.02. We also identified 80 galaxies involved in close pairs, resulting in a close pair fraction of $f_{\rm{pair}}$ = 0.09 $\pm$ 0.01. We identified 42 galaxies that were present in both samples which yielded an overlap fraction of $f_{\rm{CP+TF}}$ = 0.05 $\pm$ 0.01. The relatively small overlap fraction confirms that tidal feature samples and close pair samples detect different populations of merging galaxies, or populations in different stages of merging.
    \item We analysed the dependence of the close pair fraction and tidal feature fraction on galaxy stellar mass, and find that the fraction of galaxies in close pairs remains consistent over our mass range, whereas the fraction of galaxies with tidal features increases with increasing stellar mass. This indicates that samples relying on the presence of tidal features to detect merging galaxies are more likely to detect galaxies with higher stellar masses.
    \item We found similar colour distributions for both the tidal feature and close pair samples. Additionally, both the  tidal feature sample and close pair sample possess a larger fraction of red, and therefore more likely to be early-type, galaxies. The sample of galaxies with neither tidal features nor in close pairs, on the other hand, has an equal fraction of red and blue galaxies.
    \item We calculated the prevalence of shells, streams, plumes, and double nuclei in our sample, finding fractions $f_{\rm{shell}}$~=~0.02~$\pm$~0.01, $f_{\rm{stream}}$~=~0.11~$\pm$~0.01, $f_{\rm{plume}}$~=~0.10~$\pm$~0.01, and $f_{\rm{double~nucleus}}$~=~0.10~$\pm$~0.01 for shells, streams, plumes, and double nuclei, respectively. Shells were the least common type of tidal feature, only being present around higher mass galaxies, log$_{10}$($M_{\star}$/M$_{\odot}$)~$\geq$~10.10. Streams and plumes, on the other hand, appeared around galaxies with a wide range of colour and mass, but appeared to increase in prevalence with increasing stellar mass. Double nuclei also appeared in galaxies with a wide range of colour. 
    \item We find that galaxies with double nuclei were significantly more likely to also be classified as close pairs, with the fraction of galaxies with tidal features only possessing double nuclei being 0.28 $\pm$ 0.04 and rising to 0.93 $\pm$ 0.06 for galaxies both in close pairs and with tidal features (i.e. the CP+TF sample). This suggests that the sample of galaxies only in close pairs (CPO) likely detects early-stage mergers, where two separate galaxies are still visible, and the sample of galaxies only with tidal features (TFO) detects predominantly late-stage mergers, where only one galaxy nucleus remains visible. 
    \item The tidal feature sample is more likely to detect merging galaxies with slightly higher stellar masses. The tendency for the tidal feature sample to detect galaxies with higher stellar masses is expected if the galaxy mergers detected by this sample are predominantly late-stage mergers which have already accreted stellar mass from their merging companion. The sample of galaxies both in close pairs and with tidal features (CP+TF) appears to detect intermediate-stage mergers, with 93~±~6$\%$ of galaxies in this sample possessing visible double nuclei, compared to only 28~±~4$\%$ of galaxies in the TFO sample. Galaxies in the CP+TF sample that do not possess visible double nuclei either have a secondary galaxy too far away to detect in our postage stamp images, or the secondary galaxy is not obviously interacting.
\end{itemize}

Based on the results obtained in this paper, we recommend that future works wishing to obtain a complete picture of merging galaxies in the Universe should opt to use both close pair and tidal feature samples. When both types of samples are not available, analyses should take into account the different merging populations detected by either sample.

\section*{Acknowledgements}
\label{sec:acknowledge}
GAMA is a joint European-Australasian project based around a spectroscopic campaign using the Anglo-Australian Telescope. The GAMA input catalogue is based on data taken from the Sloan Digital Sky Survey and the UKIRT Infrared Deep Sky Survey. Complementary imaging of the GAMA regions is being obtained by a number of independent survey programmes including GALEX MIS, VST KiDS, VISTA VIKING, WISE, Herschel-ATLAS, GMRT and ASKAP providing UV to radio coverage. GAMA is funded by the STFC (UK), the ARC (Australia), the AAO, and the participating institutions. The GAMA website is \url{http://www.gama-survey.org/}.

The Hyper Suprime-Cam (HSC) collaboration includes the astronomical communities of Japan and Taiwan, and Princeton University. The HSC instrumentation and software were developed by the National Astronomical Observatory of Japan (NAOJ), the Kavli Institute for the Physics and Mathematics of the Universe (Kavli IPMU), the University of Tokyo, the High Energy Accelerator Research Organization (KEK), the Academia Sinica Institute for Astronomy and Astrophysics in Taiwan (ASIAA), and Princeton University. Funding was contributed by the FIRST program from the Japanese Cabinet Office, the Ministry of Education, Culture, Sports, Science and Technology (MEXT), the Japan Society for the Promotion of Science (JSPS), Japan Science and Technology Agency (JST), the Toray Science Foundation, NAOJ, Kavli IPMU, KEK, ASIAA, and Princeton University. 
This paper makes use of software developed for Vera C. Rubin Observatory. We thank the Rubin Observatory for making their code available as free software at http://pipelines.lsst.io/.
This paper is based on data collected at the Subaru Telescope and retrieved from the HSC data archive system, which is operated by the Subaru Telescope and Astronomy Data Center (ADC) at NAOJ. Data analysis was in part carried out with the cooperation of Center for Computational Astrophysics (CfCA), NAOJ. We are honored and grateful for the opportunity of observing the Universe from Maunakea, which has the cultural, historical and natural significance in Hawaii.\\ 
\section*{Data Availability}
\label{sec:data_avail}
 
 All data used in this work is publicly available at \url{http://www.gama-survey.org/} (for GAMA data) or at \url{https://hsc-release.mtk.nao.ac.jp/doc/} (for HSC-SSP data).


\bibliographystyle{mnras}
\bibliography{references} 





\bsp	
\label{lastpage}
\end{document}